%% file: main.tex
\PassOptionsToPackage{table,xcdraw}{xcolor}

\documentclass[pdflatex,sn-mathphys-num]{sn-jnl}

\usepackage{graphicx}
\usepackage{xcolor}
\usepackage{amssymb}
\usepackage{pifont}
\usepackage{array}
\usepackage{multirow}
\usepackage{tabularx}
\usepackage{booktabs}
\usepackage{adjustbox}
\usepackage{rotating}
\usepackage{lscape}
\usepackage{float}
\usepackage{caption}
\usepackage{subcaption}
\usepackage{enumitem}
\usepackage{soul}
\usepackage[normalem]{ulem}
\usepackage{ragged2e}
\usepackage{xspace}
\usepackage{ifthen}
\usepackage{algorithm}
\usepackage{algpseudocode}
\usepackage{listings}
\usepackage{siunitx}
\usepackage{pgf}
\usepackage{pgfplots}
\pgfplotsset{compat=1.18}
\usepackage{tikz}
\usetikzlibrary{arrows.meta, positioning, shapes.geometric, fit, backgrounds}
\usepackage[many]{tcolorbox}
\usepackage{url}
\usepackage{xurl}
\usepackage{hyperref}

\makeatletter
\def\cline#1{\@cline#1\@nil}
\makeatother

\let\table\tableorg
\let\endtable\endtableorg
\let\sidewaystable\sidewaystableorg
\let\endsidewaystable\endsidewaystableorg

\makeatletter
\renewcommand\scriptsize{\@setfontsize\scriptsize{7}{8}}
\makeatother

\soulregister\cite7
\newcolumntype{Y}{>{\raggedright\arraybackslash}X}

\newtcolorbox{mybox}[2][]{
top=0.15in,left=4pt,right=4pt,bottom=4pt,
fonttitle=\bfseries,
colbacktitle=gray,
colback=gray!5,
colframe=gray!40!black,
enhanced,
attach boxed title to top left={xshift=1.5em,yshift=-\tcboxedtitleheight/2},
boxed title style={size=small},
drop shadow={black!50!white},
title=#2,#1}

\newcounter{observation}

\newcounter{implication}

\newboolean{showcomments}
\setboolean{showcomments}{true}

\ifthenelse{\boolean{showcomments}}
{\newcommand{\nbc}[3]{
 {\colorbox{#3}{\bfseries\sffamily\scriptsize\textcolor{white}{#1}}}
 {\textcolor{#3}{\sf\small$\blacktriangleright$\textit{#2}$\blacktriangleleft$}}
 }
}
{\newcommand{\nbc}[3]{}
 \renewcommand{\hl}[1]{#1}
}

\newcommand\RQOne{Are there any differences between accepted and rejected suggested edits?}
\newcommand\RQTwo{What are the reasons for rejecting edits?}
\newcommand\RQThree{Which edits should we review?}

\graphicspath{{figures/}}

\begin{document}

\title[Understanding Suggested Edits in Stack Overflow]{Understanding Suggested Edits in Stack Overflow: A Large-Scale Empirical Study and Review Prioritization Framework}

\author*[1]{\fnm{Mehedi Hasan} \sur{Shanto}}\email{shanto1@uwindsor.ca}
\author[1]{\fnm{Muhammad} \sur{Asaduzzaman}}\email{masaduzz@uwindsor.ca}
\author[2]{\fnm{Md} \sur{Ahasanuzzaman}}\email{m\_ahasa@live.concordia.ca}
\author[1]{\fnm{Alioune} \sur{Ngom}}\email{ngom@uwindsor.ca}

\affil*[1]{\orgdiv{School of Computer Science}, \orgname{University of Windsor}, \orgaddress{\city{Windsor}, \state{Ontario}, \country{Canada}}}
\affil[2]{\orgname{Concordia University}, \orgaddress{\city{Montreal}, \state{Quebec}, \country{Canada}}}

\abstract{\input{00-abstract.tex}}

\keywords{Stack Overflow, Q\&A platforms, Suggested edits, User reputation, Community moderation, Empirical study}

\maketitle

\input{01-introduction}
\input{background.tex}
\input{02-data-collection}

\input{03a-rq1}
\input{rq2_new_v2}

\input{rq3}
\input{044-Implications}

\input{05-related-work}
\input{06-threats}

\input{07-conclusion}

\section*{Declarations}

\bmhead{Funding}
This research was supported in part by the Natural Sciences and Engineering
Research Council of Canada (NSERC).

\bmhead{Conflict of interest}
The authors declare that they have no conflict of interest.

\bmhead{Data availability}
The dataset, manual annotations, scripts, and models used in this study are
publicly available in our replication package at
\url{https://github.com/mehedisresearch/SO-Manual-Study}.

\bmhead{Ethical approval}
Not applicable. 

\bmhead{Author contributions}
Mehedi Hasan Shanto: Conceptualization, Methodology, Data collection,
Manual analysis, Software, Formal analysis, Writing -- original draft.
Muhammad Asaduzzaman: Conceptualization, Methodology, Supervision,
Writing -- review and editing.
Md Ahasanuzzaman: Manual analysis, Methodology, Writing -- review and editing.
Alioune Ngom: Supervision, Writing -- review and editing.
All authors read and approved the final manuscript.

\bibliography{references}

\begin{appendices}
\input{09-appendix}

\end{appendices}

\end{document}

%% file: 00-abstract.tex
Suggested edits allow low-reputation users to improve Stack Overflow posts and contribute to the platform’s knowledge base. However, we still know little about why these edits are accepted or rejected and how reviewers can efficiently identify problematic edits. In this study, we analyze 3,888,805 suggested edits submitted between 2011 and May 2023. First, we compare accepted and rejected edits using 35 features. We find significant differences between them, with the largest differences related to users’ editing history. Second, we manually analyze rejected edits and identify 17 fine-grained rejection reasons. These reasons provide a more detailed understanding of why edits are rejected than Stack Overflow’s predefined rejection reasons. Third, we evaluate three models for prioritizing edits based on their likelihood of rejection. The Feature Model, built using the 35 studied features, provides the best trade-off between review rate and rejected-edit recall. At a decision threshold of $\tau=0.40$, it prioritizes 52.3\% of edits for human review while capturing 80.8\% of rejected edits. In comparison, the Embedding Model captures 78.8\% of rejected edits while prioritizing 60.9\% of edits, and the EditEx Model captures 72.5\% while prioritizing 59.0\%. We also conduct a complementary zero-shot evaluation on a separate set of 500 edits. The Feature Model achieves 78.0\% recall at $\tau=0.40$, compared with 23.6\%--31.2\% for GPT-4o, Qwen2.5-32B, and Llama-3.3-70B. Overall, our findings provide a detailed understanding of suggested-edit rejection and show that rejection-risk prediction can help prioritize potentially problematic edits for human review.

%% file: 01-introduction.tex
\section{Introduction}

Stack Overflow is one of the largest and most popular community question-and-answer platforms for discussing programming-related questions. Millions of developers use Stack Overflow to seek solutions, share technical knowledge, and learn from others' experiences. To maintain the quality and usefulness of this continuously evolving knowledge base, Stack Overflow relies heavily on community editing. A particularly important mechanism is the \emph{suggested edit}, which enables users with limited reputation to contribute improvements to existing questions and answers. Users can suggest changes such as correcting errors, clarifying explanations, improving formatting, and updating code snippets. However, unlike regular edits made by users with sufficient reputation, suggested edits are not applied immediately. In general, users with fewer than 2,000 reputation points can propose \emph{suggested edits}, which are placed in a peer-review queue and must be approved by eligible community reviewers before being applied to the post. In contrast, users who have earned the editing privilege can make \emph{direct edits} to others' posts without review \cite{so-edit-policy}. Figure~\ref{fig:edit_process_diagram} illustrates Stack Overflow's suggested-edit review process.

Suggested edits therefore serve a broader role than content maintenance alone. They provide an accessible pathway through which low-reputation and relatively new users can begin contributing to the community without first having to ask or answer questions. An accepted suggested edit also earns the contributor reputation, allowing editing activity to contribute toward gaining additional community privileges \cite{so_edit_reward}. More broadly, the review process exposes contributors to community expectations about what constitutes a useful edit: reviewers either approve the proposed contribution or reject it. Suggested editing can therefore serve simultaneously as a mechanism for improving Stack Overflow's knowledge base, attracting low-reputation users to contribute, and helping newcomers gain experience and reputation as they become more involved in the community.

This model, however, places substantial responsibility on reviewers. Every suggested edit must be examined even when the proposed change is trivial, unnecessary, incorrect, or otherwise unlikely to improve the post. Reviewers must therefore assess a large volume of submissions while identifying and filtering problematic edits among legitimate contributions to maintain the quality of Stack Overflow content. This concern has also been expressed within the Stack Overflow community. For example, one reviewer stated in a Meta Stack Overflow discussion\footnote{https://meta.stackoverflow.com/questions/393396/should-we-allow-low-rep-question-owners-to-unilaterally-accept-edits}: \textbf{\emph{``I would rather look at preventing bad edits than require more users to review good edits.''}} Community members have similarly raised concerns about trivial edits being approved\footnote{https://meta.stackoverflow.com/questions/381687/}. Such decisions matter beyond reviewer workload: inappropriate approvals may affect post quality, while unclear or inconsistent rejections may provide limited guidance to contributors who are still learning community editing norms. Repeated rejections without sufficiently informative feedback may discourage further participation and hinder their progression toward becoming active community contributors. Therefore, a better understanding of why suggested edits are rejected, and whether likely rejected edits can be identified in advance, could help reviewers prioritize their efforts and provide more useful guidance to contributors.

Despite the importance of the suggested-edit review process, prior studies have mainly examined edits that were already applied to Stack Overflow posts and studied how such edits affect content evolution and quality. Zhang et al.~\cite{obsoluteanswer} studied outdated answers, Baltes et al.~\cite{baltes2018sotorrent} analyzed post edit histories, Jin and Servant~\cite{jin2019editsHighlyAnswered} examined the relationship between editing and the number of answers, and Wang et al.~\cite{wang2018reviseAnswers} investigated answer revisions over time. The closest work, Mondal et al.~\cite{autopredsaikat1}, studied \emph{rollback edits}, i.e., edits that were first applied to a post but later reverted to a previous version, whereas rejected suggested edits are stopped during peer review and never become part of the post. Thus, prior work does not explicitly examine \emph{suggested edits}, the characteristics associated with their rejection, the reasons they are rejected, or how such information can be used to support the review process.

To address this gap, we conduct a large-scale empirical study of suggested edits on Stack Overflow. We construct and analyze a dataset of 3,888,805 suggested edits collected up to May 2023. We first conduct a quantitative analysis to examine how rejected suggested edits differ from accepted ones. We then perform a qualitative analysis to understand the reasons and patterns associated with rejection. Finally, we evaluate a technique for identifying edits that are likely to be rejected and thereby supporting the review process. Specifically, we study the following three research questions.

\begin{enumerate}

\item {\textbf{RQ1: \RQOne}}\newline
To compare accepted and rejected suggested edits, we extract 35 features 
spanning five dimensions: post-content, post history, change 
metrics, readability, and user history. We then compare accepted and rejected edits using the Mann--Whitney U test with Bonferroni correction \cite{abdi2007bonferroni} and Cliff's $\delta$ effect size.\par 
\vspace{0.1cm}
\textit{Results: Our analysis shows that accepted and rejected edits are distinguishable, with the strongest differences appearing in the user-history dimension. In particular, prior acceptance rate shows a medium effect size, suggesting that a user's editing history is a stronger signal of rejection risk than any individual structural characteristic of the edit. Rejected edits also tend to target posts with higher scores and more upvotes, and to involve larger textual changes, although these differences are modest. In contrast, readability changes show little practical distinction between accepted and rejected edits.}
\item {\textbf{RQ2: \RQTwo}}\newline
To understand why suggested edits are rejected, we conduct a manual coding study of 690 randomly sampled rejected edits, selected with a 99\% confidence level and 5\% margin of error. We first use open coding on an initial subset of 100 edits to develop the coding scheme. The first two authors then independently code all 690 edits, achieving substantial agreement ($\kappa=0.80$), and resolve disagreements through discussion, with a third author involved when needed. \par
\vspace{0.1cm}
\textit{Results: We identify 17 fine-grained rejection reasons that refine Stack Overflow's broad predefined rejection reasons and make several implicit rejection patterns explicit. Minor formatting changes are the most common reason for rejection (35.94\%), and more than half of the rejected edits (52.61\%) involve changes that provide little substantive improvement. We also find that nearly one-fifth (19.86\%) misuse the edit mechanism to ask questions, add comments, provide alternate suggestions, or post status updates. Notably 25.51\% of the rejected edits introduce incorrect and harmful changes that can degrade post correctness, usefulness, or integrity, highlighting the importance of identifying such edits for careful human review.
}
\item {\textbf{RQ3: \RQThree}}\newline
To prioritize suggested edits for review, we compare three techniques for predicting rejected edits: an EditEx-based model~\citep{autopredsaikat1}, an embedding-based model, and a feature-based model built from the characteristics studied in RQ1. Each model assigns a rejection probability to a suggested edit, which we use to rank edits for review. At threshold $\tau$, edits with scores at or above $\tau$ are considered to be problematic (i.e., likely to be rejected) and therefore they are sent for human review. The remaining edits can wait in the regular review queue, but platforms may choose to automatically accept them after a configurable waiting period if no reviewer intervenes and the estimated risk is sufficiently low. We measure performance  using two metrics: \emph{Review rate}, the proportion of submitted edits prioritized for human review, and \emph{recall}, the proportion of rejected edits correctly identified. Together, \emph{review rate} measures the human effort required, while \emph{recall} measures the proportion of rejected edits detected, thereby capturing the trade-off between reviewer workload and detection effectiveness. \par
\vspace{0.1cm}
\textit{Results: The feature model achieves the best balance between the proportion of edits prioritized for review and the proportion of rejected edits identified. For example, at $\tau=0.40$, it prioritizes 52.3\% of edits for human review while capturing 80.8\% of rejected edits, outperforming both the embedding and EditEx models. Even under stricter prioritization, the model remains effective. At $\tau=0.80$, the model prioritizes only 14.6\% of edits while still capturing  41.7\% of rejected edits. These results show that the feature model can effectively prioritize potentially problematic edits while reducing reviewer workload.}

\end{enumerate}


Our findings offer a deeper understanding of how suggested edits are 
reviewed and rejected, and how moderation mechanisms can be improved. 
Although our study focuses on Stack Overflow, the implications extend to 
other collaborative knowledge platforms that rely on human review to 
maintain content quality. We make the following concrete contributions:

\begin{itemize}
    \item \textbf{Large-scale empirical characterization.} We present the largest empirical study of suggested edits to date, analyzing 3,888,805 suggested edits submitted between 2011 and May 2023 under their 
    natural class imbalance, identifying statistically significant 
    differences between accepted and rejected edits across 35 features 
    and five dimensions.

    \item \textbf{Refined rejection taxonomy.} Through manual coding of 
    a random sample with high inter-rater agreement ($\kappa = 0.80$), 
    we derive a taxonomy of 17 fine-grained rejection reasons with concrete 
    examples, replacing the platform's broad and ambiguous predefined labels 
    and providing actionable guidance for reviewers and contributors.

    \item \textbf{Threshold-based review prioritization framework.} We 
    propose and evaluate a probability-based prioritization framework that 
    ranks suggested edits by rejection risk, enabling reviewers to focus 
    effort on the most problematic submissions. Our feature-based model provides a more favorable review-rate/recall trade-off at selected operating points, particularly $\tau = 0.40$,and outperforms three state-of-the-art zero-shot LLMs 
    (GPT-4o, Qwen2.5-32B, Llama-3.3-70B) on rejected-class detection.

\end{itemize}

The remainder of this paper is organized as follows. Section~2 provides the needed background to understand the review mechanism and suggested edits. Section~3 introduces our studied dataset. The study procedures and the results of our three research questions are described in sections 4, 5, and 6, respectively. We discuss the outcome of the study and describe prior work related to our study in Section 7 and 8. Section 9 presents threats to the validity of our study. Finally, Section 10 concludes our study. Our study materials and full replication package are available at 
GitHub~\cite{replication}.

%% file: background.tex
\section{Background}
\begin{figure}[ht]
    \centering

    \includegraphics[scale=0.3]
    {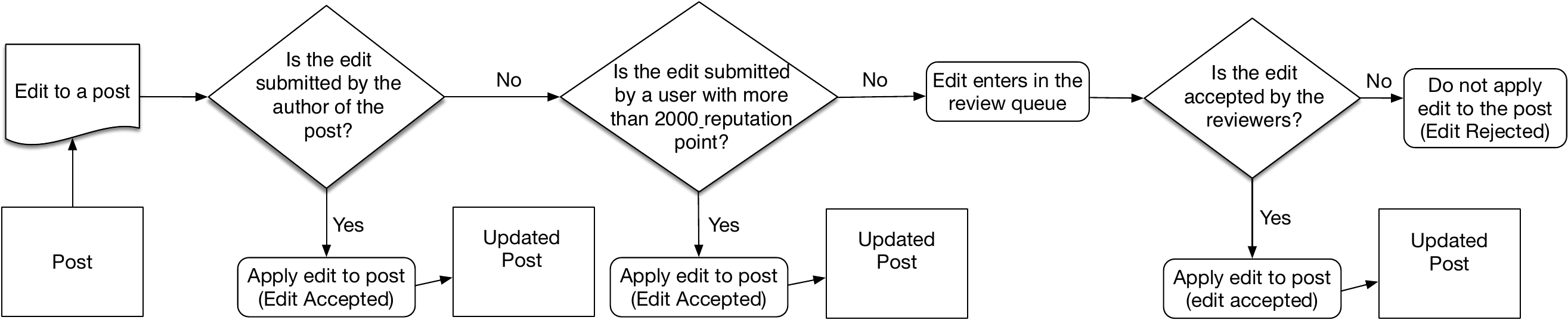}
    \caption{The process of suggested-edit review process on Stack Overflow.} 
    

    \label{fig:edit_process_diagram}
\end{figure}

This section provides an overview of the editing process in Stack Overflow. Specifically, we describe the question-and-answer process, how users revise their posts (either questions or answers) (Figure \ref{fig:question_answering_process}), and how the suggested-edit review mechanism operates in Stack Overflow  (Figure \ref{fig:edit_process_diagram}).

\subsection{Question and answer process in Stack Overflow }
\begin{figure}[ht]
    \centering
    \includegraphics[width=0.9\linewidth]{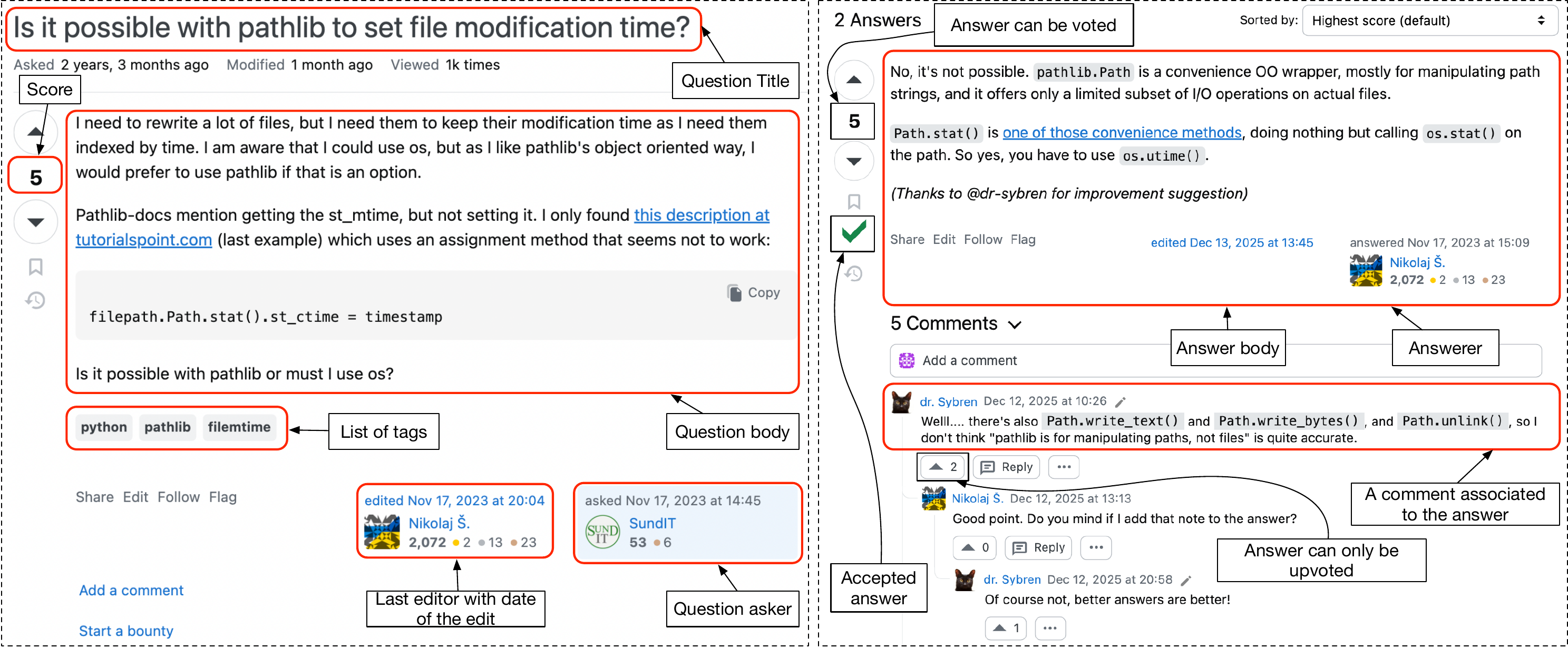}
    \caption{An example of a question (left) and the accepted answer (right)}
    \label{fig:question_answering_process}
\end{figure}

To understand how the editing mechanism operates on Stack Overflow, we first briefly describe the structure of its question-and-answer process. On Stack Overflow, registered users can ask programming-related questions, each consisting of a title, body, and a set of tags that indicate its topic. Other users can provide answers, and the author of the question can mark one answer as accepted. Both questions and answers can receive comments for clarification and discussion, as well as upvotes and downvotes from the community. The difference between the numbers of upvotes and downvotes determines a post's score, which reflects the community's assessment of its usefulness and quality. Figure~\ref{fig:question_answering_process} shows an example of a question and its accepted answer. 


\subsection{Editing process in Stack Overflow }

\noindent{\textbf{Direct edits.}} Stack Overflow  allows users to edit questions and answers to improve their clarity, correctness, formatting, and keep their content up to date. Post authors can directly edit their own posts regardless of their reputation, and these changes are applied immediately without peer review. Users with at least 2,000 reputation points have the editing privilege and can \textit{directly edit} other users' posts without peer review. These edits are applied immediately and do not require validation or approval from other community members~\cite{so-edit-policy}. \vspace{0.1cm}

\noindent{\textbf{Suggested edits and the review queue.}}
Users who have not earned the editing privilege can still propose changes to other users' posts through \emph{suggested edits}. Even users with zero reputation can submit suggested edits, providing an accessible entry point for newcomers to contribute to Stack Overflow. Unlike direct edits, suggested edits are not applied immediately; instead, they are placed in the Suggested Edits review queue, where eligible community members evaluate the proposed changes~\cite{so-edit-policy}. Reviewers can approve, reject or improve a suggested edit. In general, two approval votes apply the edit to the post, whereas two rejection votes discard it and leave the post unchanged. If a reviewer considers the edit useful but incomplete, the reviewer may further improve it before applying the changes. Figure~\ref{fig:edit_process_diagram} illustrates this review process. 


\begin{figure}[ht]
    \centering
    \includegraphics[scale=0.27]{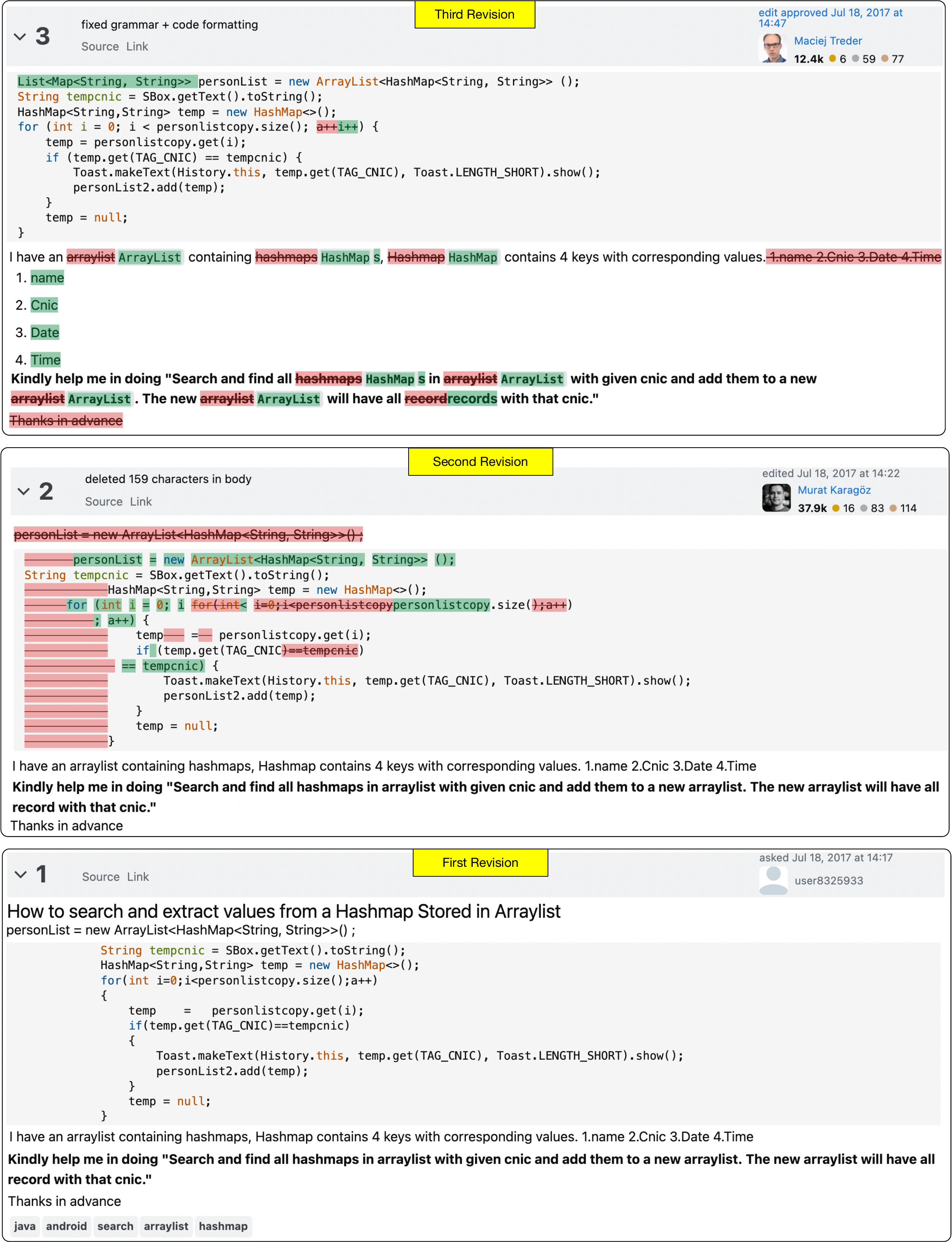}
    \caption{Revision history of a Stack Overflow question.}
    \label{fig:revision_history}
    \vspace{-5mm}
\end{figure}

%% file: 02-data-collection.tex
\section{Suggested-Edit Dataset Construction}
\label{sec:data_collection}

A key contribution of our study is the construction of a large-scale dataset of suggested edits and their review outcomes on Stack Overflow. In this section, we describe the process used to construct this suggested edit dataset.

\smallskip\noindent\textbf{Data sources.} The Stack Overflow  data dump contains user-generated content in multiple XML files, including users, posts, comments, post histories, links, tags, and votes. However, the Stack Overflow  data dump does not contain suggested edits or their corresponding review decisions. To obtain suggested-edit information, we used the Stack Exchange Data Explorer (SEDE), a platform that allows users to run SQL queries on the Stack Overflow  dataset hosted online. This platform provides access to additional data beyond that available in the Stack Overflow  data dump, structured as a set of relational tables. Specifically, we query four tables that are relevant for our study: \texttt{SuggestedEdits}, \texttt{ReviewTasks}, \texttt{ReviewTaskResults}, and \texttt{ReviewRejectionReasons}. The \texttt{SuggestedEdits} table stores the proposed edits submitted by users (e.g., users with reputation lower than 2000), including the post-id, user-id, edit content fields, and timestamps indicating whether the edit was approved, rejected, or still pending. A suggested edit is classified as \emph{accepted} when its \texttt{ApprovalDate} is non-null and as \emph{rejected} when its \texttt{RejectionDate} is non-null. Each suggested edit generates a corresponding entry in the \texttt{ReviewTasks} table, which represents items placed into community review queues and records the task type, creation date, active/completed state, and the link back to the originating suggested edit. When reviewers take action, the result is captured in \texttt{ReviewTaskResults}, which logs the specific outcome (e.g., Approve, Reject), the review date, any comment provided, and, for rejections, the associated reason code. Those reason codes are defined in \texttt{ReviewRejectionReasons}, a lookup table containing the predefined list of reasons reviewers may select when rejecting an edit, ensuring consistency and interpretability in moderation decisions.

All information related to posts and users is obtained from the May 2023 Stack Overflow  data dump. This includes the content and revision history of posts, user information, comments, and voting records associated with each post. We use these data to reconstruct the state and history of the corresponding post and contributor at the time the suggested edit was submitted.

\smallskip\noindent\textbf{Data retrieval and post linking.} The SEDE limits each query result to 50,000 rows. We therefore partition the data and retrieve the records through multiple SQL queries. We then link each suggested edit to its corresponding post using \texttt{Posts.xml} from the Stack Overflow  data dump. Some suggested-edit records refer to posts that were subsequently deleted and are no longer available in the data dump. We remove these records because their corresponding post content cannot be recovered.

\smallskip\noindent\textbf{Filtering suggested edits.} We exclude suggested edits that are rejected due to \emph{suggested-edit conflicts}. These rejections occur because of concurrent editing rather than reviewers' assessment of the proposed change. Including them could therefore conflate procedural conflicts with rejections related to the content or quality of an edit. Therefore, we remove 147,453 suggested edits that are rejected due to \emph{suggested-edit conflicts}.

\smallskip\noindent 
In summary, we initially collected 4,036,258 suggested edits submitted up to May 2023, comprising 3,279,310 accepted edits and 756,948 rejected edits. Among the rejected edits, 147,453 were rejected because of suggested-edit conflicts and were excluded from subsequent analyses. The final dataset therefore contains 3,888,805 suggested edits: 3,279,310 accepted edits and 609,495 non-conflict rejected edits. The complete scripts for data retrieval and processing are available in our replication package~\cite{replication}.

%% file: 03a-rq1.tex
\section{RQ1: \RQOne} 
\label{rq_one}

\subsection{Motivation}
Understanding the differences between accepted and rejected suggested edits helps us determine whether rejection decisions follow consistent and measurable patterns. It may depend on structural properties of the edit, characteristics of the post, or the history of the contributing user. We are interested in learning whether these two groups differ not only in terms of change size and readability impact, but also in terms of post context and contributor behavior. Investigating these differences provides a deeper understanding of how community moderation operates on Stack Overflow. It also allows us to identify objective signals associated with accepted and rejected review outcomes. These findings form the foundation for building predictive models in later research questions and help us move from descriptive analysis toward actionable review optimization.

\subsection{Approach}

To answer RQ1, we follow three steps. First, we reconstruct the post revisions associated with each suggested edit to identify the state of the post immediately before the edit (i.e., pre-edit) and the version proposed by the editor (i.e., post-edit) (Section~\ref{sec:revision_reconstruction}). Second, using the reconstructed versions and the associated post and user information, we extract 35 features across five dimensions: post content, post history, change metrics, readability, and user history (Section~\ref{sec:feature_collection}). Finally, we group the suggested edits into accepted and rejected edits based on their recorded review outcomes and compare the two groups to identify which features differ significantly and assess the magnitude of these differences (Section~\ref{sec:rq1_evaluation}).

\subsubsection{Reconstruction of Post Revisions}
\label{sec:revision_reconstruction}

A single post can receive multiple suggested edits over time. Therefore, for each suggested edit, we reconstruct the corresponding pre-edit and proposed post-edit versions of the post. The resulting pre-edit and post-edit versions of a post allow us to measure the changes introduced by that specific suggested edit, rather than changes accumulated across multiple post revisions. To reconstruct the pre-edit version of a suggested edit, we use the \texttt{PostHistory} table and identify the most recent revision of the post before the suggested edit's \texttt{CreationDate}. This revision represents the state of the post when the suggested edit was submitted. The reconstruction of the proposed post-edit version differs for accepted and rejected edits. For an accepted edit, the proposed changes are applied to the post and recorded as a new revision in the \texttt{PostHistory} table. We use this revision to obtain and validate the post-edit version. For a rejected edit, the proposed changes are not applied and therefore no corresponding \texttt{PostHistory} revision is created. In this case, we reconstruct the proposed post-edit version using the title, body, and tag fields stored in the \texttt{SuggestedEdits} table.

By pairing the pre-edit and proposed post-edit versions for each suggested edit, we isolate the changes introduced by that edit and use them to compute change-based features such as text difference, code modification, and structural changes described in the following sections.

\subsubsection{Feature Collection}
\label{sec:feature_collection}

We collect 35 features across five dimensions to compare accepted and rejected suggested edits: post content, post history, change metrics, readability, and user history. Table~\ref{tab:feature_change_post} summarizes the features in each dimension, along with their descriptions and rationales.

\subsubsection*{Post-Content}
Post-content features characterize the post associated with a suggested edit. Prior studies on Stack Overflow  suggest that characteristics of a post (e.g., popularity) can influence how it is perceived and moderated by the community \cite{anderson2012discovering}. Therefore, we compute five features from the pre-edit version of each post: post length, number of tags, tag popularity, number of links, and number of code lines. Post length is measured using the number of characters in the post body. Tag count is derived by parsing the tag string and counting the number of tag entries. Tag popularity is measured using the frequency of each tag
among posts created before the submission of the
corresponding suggested edit. We identify code blocks based on Stack Overflow's code-formatting convention, specifically the \texttt{<pre><code>...</code></pre>} markup \cite{baltes2018sotorrent}, and count the number of lines contained within these blocks.

\subsubsection*{Post-History}
Post-history features capture the state and prior activity of the post before the suggested edit was submitted. Using the \texttt{PostHistory} and \texttt{SuggestedEdits} tables, we compute the number of previous edits, comments, previously accepted suggested edits, and previously rejected suggested edits associated with the post. We also obtain the post's upvotes, downvotes, and score from the \texttt{Votes} table. For all historical features, we consider only records created before the current suggested edit. Studying these features helps us understand whether community visibility and post maturity influence edit outcome decisions \cite{autopredsaikat1}.

\subsubsection*{Change Metrics}

Change metrics quantify the modifications introduced by a suggested edit. We compare the reconstructed pre-edit and proposed post-edit versions to measure changes to the title, body, tags, links, code, and overall content. We also compute a normalized text-change ratio. The normalized text change ratio is computed as the number of 
characters added or removed in the plain-text portion of the post 
(excluding code blocks) divided by the total number of characters 
in the plain-text portion of the original post. Code segments are separated from plain text using Markdown formatting so that code changes can be measured separately \cite{autopredsaikat1}. In addition, we compute the delay between post creation and suggested-edit submission. This feature captures whether edits were made shortly after posting or long after the post became stable. Prior research suggests that edits made early in a post’s lifecycle may be evaluated differently than late-stage edits \cite{mamykina2011design}.

\subsubsection*{Readability}

 Readability features characterize changes in the linguistic complexity of a post. Prior research on Stack Overflow has examined the linguistic characteristics of technical answers and their relationship with users' preferences \cite{Kabir_2024}. For each suggested edit, we extract the plain-text body from the pre-edit and proposed post-edit versions after removing code blocks, HTML tags, and formatting artifacts. We compute eight readability measures: Flesch Reading Ease, Gunning Fog Index, LIX, Flesch--Kincaid Grade Level, Automated Readability Index, Coleman--Liau Index, SMOG Grade, and Dale--Chall Score, following prior work \cite{arpon}. We calculate these measures using the Python \texttt{readability} library\footnote{https://pypi.org/project/readability/}. For each measure, we use the difference between the proposed post-edit and pre-edit scores to represent the readability change introduced by the suggested edit.

\subsubsection*{User-History.}
User-history features characterize a contributor's editing activity and experience before submitting a suggested edit. Prior studies suggest that contributor history may be associated with how edits are evaluated by the community \cite{rollbackincnsitency}. Using the \texttt{SuggestedEdits} table, we calculate the number of previously accepted, rejected, and total suggested edits submitted by each contributor, along with their historical acceptance and rejection rates. These rates are computed by dividing the number of previously accepted or rejected edits by the total number of suggested edits submitted before the current edit. For contributors without prior suggested edits, both rates are set to zero to indicate the absence of editing history. We also measure short-term editing activity by counting the suggested edits submitted during the preceding 24 hours. We estimate the contributor's reputation at the time of each suggested edit by aggregating reputation-changing events from the \texttt{Votes} and \texttt{Posts} tables, following prior work \cite{autopredsaikat1}. This reconstruction is necessary because the Stack Overflow data dump provides reputation values at the time of the data snapshot rather than at individual edit submission times. To avoid temporal leakage, we consider only events recorded before the submission of the corresponding suggested edit. Similarly, all other user-history features are calculated using information available before the current edit.

\newcolumntype{Y}{>{\raggedright\arraybackslash}X}                  
\newcolumntype{Z}{>{\raggedright\arraybackslash\scriptsize}X}     

\begin{table*}[!]
\centering
\caption{Studied features in five dimensions.}
\label{tab:feature_change_post}
\tiny
\setlength{\tabcolsep}{2pt}
\renewcommand{\arraystretch}{1.15}

\begin{tabularx}{\linewidth}{|>{\raggedright\arraybackslash}m{1.7cm}|l|Y|Z|}
\hline
\textbf{Dimension} & \textbf{Feature} & \textbf{Description} & \textbf{Rationale} \\
\hline

\multirow{5}{*}{\rotatebox{0}{Post-Content}}
& PostLength & Characters count in the post body &
\multirow{5}{=}{Capture size, popularity, and engagement of posts; influence reviewer attention and leniency.} \\
\cline{2-3}
& NumTags & Number of tags in the question post & \\
\cline{2-3}
& TagPopularity & Aggregate popularity of tags & \\
\cline{2-3}
& NumLinks & Hyperlinks in body & \\
\cline{2-3}
& NumCodeline & Lines of code in the post  & \\ 

\noalign{\vskip 4pt}\hline

\multirow{7}{*}{\rotatebox{0}{Post-History}}
& NumEditsBefore & Previous edits on the post &
\multirow{7}{=}{Capture historical evolution of the post} \\
\cline{2-3}
& CommentCount & Comments on post  & \\
\cline{2-3}
& PostAcceptedEditsBefore & Accepted edits in the post before  & \\
\cline{2-3}
& PostRejectedEditsBefore & Rejected edits in the post before  & \\ \cline{2-3}
& Score & Net votes & \\
\cline{2-3}
& Upvotes & Upvotes & \\
\cline{2-3}
& Downvotes & Downvotes & \\
\noalign{\vskip 4pt}\hline

\multirow{8}{*}{\rotatebox{0}{Change Metrics}}
& TitleLengthChanged & Characters changed in the title of a question post &
\multirow{8}{=}{Measure magnitude of edits; large or drastic edits may reduce acceptance, small precise ones may increase it.} \\
\cline{2-3}
& BodyLengthChanged & Characters changed in body & \\
\cline{2-3}
& TagLengthChanged & Number of tag changed in the question post & \\
\cline{2-3}
& TotalLengthChanged & Total chars changed & \\
\cline{2-3}
& NumLinksChanged & Hyperlinks modified & \\
\cline{2-3}
& CodelineChanged & Lines of code changed & \\
\cline{2-3}
& TextChangeRatio & Normalized text change ratio & \\
\cline{2-3}
& Delay (hours) & Delay between creation and edit & \\
\noalign{\vskip 4pt}\hline

\multirow{8}{*}{\rotatebox{0}{Readability}}
& Flesch & Reading Ease score &
\multirow{8}{=}{Approximate linguistic complexity; harder text may hinder comprehension and reduce acceptance. We compute the delta of readability score using previous and edited body of suggested edits.} \\
\cline{2-3}
& Fog & Gunning Fog index & \\
\cline{2-3}
& LIX & LIX index & \\
\cline{2-3}
& Kincaid & Flesch--Kincaid grade & \\
\cline{2-3}
& ARI & Auto Readability Index  & \\
\cline{2-3}
& ColemanLiau & Coleman--Liau index & \\
\cline{2-3}
& SMOG & SMOG grade & \\
\cline{2-3}
& DaleChall & New Dale--Chall score. & \\
\noalign{\vskip 4pt}\hline

\multirow{7}{*}{\rotatebox{0}{User-History}}
& Reputation & Estimated contributor reputation at edit submission &
\multirow{7}{=}{Capture contributor experience and historical editing activity to examine their association with review outcomes.} \\
\cline{2-3}
& UserAcceptedEditsBefore & Prior accepted suggested edits by the contributor & \\
\cline{2-3}
& UserRejectedEditsBefore & Prior rejected suggested edits by the contributor & \\
\cline{2-3}
& UserTotalEditsBefore & Total prior suggested edits by the contributor & \\
\cline{2-3}
& UserAcceptanceRate & Proportion of prior suggested edits accepted & \\
\cline{2-3}
& UserRejectionRate & Proportion of prior suggested edits rejected & \\
\cline{2-3}
& UserEdits24h & Suggested edits submitted in the preceding 24 hours & \\
\hline

\end{tabularx}

\vspace{2pt}
\parbox{\linewidth}{\scriptsize
\textit{Note:} UserAcceptanceRate and UserRejectionRate
are set to zero for contributors with no prior suggested edits.
}
\vspace{-0.5cm}
\end{table*}

\subsubsection{Statistical Comparison between accepted and rejected edits}
\label{sec:rq1_evaluation}

The unit of analysis is an individual suggested edit. After
extracting the features for each edit, we divide the edits
into accepted and rejected groups based on their recorded
review outcomes. For each feature, we summarize its
distribution within each group using the mean, median,
minimum, and maximum values.

We compare the two groups using the Mann--Whitney U test
\cite{mann}. We use this non-parametric test because the
feature distributions can be skewed and are not assumed
to follow a normal distribution. Since we perform a
separate statistical test for each feature, the probability
of obtaining a significant result by chance increases
with the number of tests. To mitigate this issue, we
apply the Bonferroni correction and consider a difference
statistically significant when the adjusted p-value is
below 0.05 \cite{abdi2007bonferroni}. To assess the magnitude of the differences, we report
Cliff's delta ($\delta$), a non-parametric effect size
aligned with the Mann--Whitney U test \cite{cliff,romano}.
Positive $\delta$ values indicate that accepted edits
tend to have larger feature values than rejected edits,
while negative values indicate the opposite. We classify
effect sizes as negligible ($|\delta| < 0.147$),
small ($0.147 \leq |\delta| < 0.330$),
medium ($0.330 \leq |\delta| < 0.474$),
or large ($|\delta| \geq 0.474$), following
\citet{romano}. This analysis allows us to identify
which features differ significantly between accepted
and rejected edits and whether those differences are
practically meaningful.

\newcommand{\pv}[1]{%
  \pgfmathparse{#1 < 0.05}%
  \ifnum\pgfmathresult=1
    $\leq$0.05 %
  \else
    $\geq$0.05 %
  \fi
}

\begin{table*}[!]
\centering
\caption{The studied numeric features (aligned to the feature set in Table~\ref{tab:feature_change_post}), with p-values for significance between accepted and rejected edits and Cliff's $\delta$ as the non-parametric effect size.}
\label{tab:pvalues_featureset}
\scriptsize
\begin{adjustbox}{max width=\textwidth}
\begin{tabular}{|l|l|c|c|c|}
\hline
\multirow{2}{*}{\textbf{Dimension}} &
\multirow{2}{*}{\textbf{Feature}} &
\multicolumn{2}{c|}{\textbf{Mean/Median/Min/Max}} &
\multirow{2}{*}{Adjusted \textbf{P-value \& $|\delta|$}} \\
\cline{3-4}
 & & \textbf{Accepted} & \textbf{Rejected} & \\
\hline

\multirow{5}{*}{Post-Content}
& PostLength & 1329.04 / 805 / 28 / 30000 & 1230.41 / 760 / 27 / 30000 & \pv{0.000000e+00} \& 0.04 (N) \\ \cline{2-5}
& NumTags & 2.07 / 2 / 0 / 6 & 1.39 / 1 / 0 / 5 & \pv{0.000000e+00} \& 0.24 (S) \\ \cline{2-5}
& TagPopularity & 587749.54 / 241069 / 0 / 7670705 & 445486.70 / 24706 / 0 / 6751210 & \pv{0.000000e+00} \& 0.19 (S) \\ \cline{2-5}
& NumLinks & 0.480 / 0 / 0 / 386 & 0.543 / 0 / 0 / 376 & \pv{0.000000e+00} \& 0.02 (N) \\ \cline{2-5}
& NumCodeline & 1.94 / 0 / 0 / 1514 & 1.38 / 0 / 0 / 880 & \pv{7.329134e-281} \& 0.04 (N) \\ \cline{2-5}
\hline

\multirow{7}{*}{Post-History}
& NumEditsBefore & 0.109 / 0 / 0 / 65 & 0.225 / 0 / 0 / 138 & \pv{0.000000e+00} \& 0.04 (N) \\ \cline{2-5}
& CommentCount & 0.040 / 0 / 0 / 29 & 0.087 / 0 / 0 / 50 & \pv{0.000000e+00} \& 0.01 (N) \\ \cline{2-5}
& PostAcceptedEditsBefore & 0.899 / 1 / 0 / 56 & 0.877 / 1 / 0 / 57 & \pv{0.000000e+00} \& 0.02 (N) \\ \cline{2-5}
& PostRejectedEditsBefore & 0.031 / 0 / 0 / 46 & 0.124 / 0 / 0 / 138 & \pv{0.000000e+00} \& 0.03 (N) \\ \cline{2-5}
& Score & 10.63 / 1 / -67 / 33053 & 40.29 / 2 / -204 / 34455 & \pv{0.000000e+00} \& 0.20 (S) \\ \cline{2-5}
& Upvotes & 11.14 / 1 / 0 / 33080 & 40.94 / 2 / 0 / 34486 & \pv{0.000000e+00} \& 0.20 (S) \\ \cline{2-5}
& Downvotes & 0.516 / 0 / 0 / 190 & 0.650 / 0 / 0 / 280 & \pv{0.00000} \& 0.00 (N) \\ \cline{2-5}
\hline

\multirow{8}{*}{Change Metrics}
& TitleLengthChanged & 0.097 / 0 / 0 / 128 & 0.084 / 0 / 0 / 125 & \pv{1.201802e-96} \& 0.04 (N) \\ \cline{2-5}
& BodyLengthChanged & 16.39 / 4 / 0 / 29850 & 46.43 / 8 / 0 / 29298 & \pv{0.000000e+00} \& 0.16 (S) \\ \cline{2-5}
& TagLengthChanged & 0.685 / 0 / 0 / 7 & 0.398 / 0 / 0 / 7 & \pv{0.000000e+00} \& 0.03 (N) \\ \cline{2-5}
& TotalLengthChanged & 69.48 / 16 / 0 / 29850 & 140.75 / 27 / 0 / 29504 & \pv{0.000000e+00} \& 0.15 (S) \\ \cline{2-5}
& NumLinksChanged & 0.234 / 0 / 0 / 400 & 0.211 / 0 / 0 / 382 & \pv{4.425266e-24} \& 0.04 (N) \\ \cline{2-5}
& CodelineChanged & 0.688 / 0 / 0 / 1176 & 0.461 / 0 / 0 / 1003 & \pv{2.667174e-245} \& 0.08 (N) \\ \cline{2-5}
& TextChangeRatio & 0.095 / 0.039 / 0 / 1 & 0.135 / 0.052 / 0 / 1 & \pv{0.000000e+00} \& 0.06 (N) \\ \cline{2-5}
& Delay (hours) & 5009.31 / 0.486 / 0 / 128056.95 & 8048.44 / 1.80 / 0 / 127697.79 & \pv{0.000000e+00} \& 0.09 (N) \\ \cline{2-5}
\hline

\multirow{8}{*}{Readability}
& Flesch & 2.41 / 0.180 / -32.55 / 48.72 & 1.13 / 0 / -28.91 / 41.03 & \pv{0.000000e+00} \& 0.01 (N) \\ \cline{2-5}
& Fog &  0.160 / 0 / -4.80 / 3.92 & 0.10 / 0 / -4.37 / 3.88 & \pv{0.000000e+00} \& 0.01 (N) \\ \cline{2-5}
& LIX & 0.910 / -0.010 / -16.27 / 14.83 & 0.740 / 0 / -15.90 / 15.22 & \pv{0.000000e+00} \& 0.02 (N) \\ \cline{2-5}
& Kincaid & 0.070 / 0 / -6.44 / 6.08 & 0.110 / 0 / -6.01 / 7.14 & \pv{0.000000e+00} \& 0.01 (N) \\ \cline{2-5}
& ARI & 0.040 / 0 / -9.32 / 9.87 & 0.080 / 0 / -9.07 / 11.21 & \pv{0.000000e+00} \& 0.01 (N)\\ \cline{2-5}
& ColemanLiau & 0.010 / 0 / -8.44 / 7.56 & 0.030 / 0 / -7.90 / 9.13 & \pv{0.000000e+00} \& 0.01 (N)\\ \cline{2-5}
& SMOG & 0.030 / 0 / -2.32 / 2.51 & 0.040 / 0 / -2.11 / 2.93 & \pv{0.000000e+00} \& 0.00 (N)\\ \cline{2-5}
& DaleChall & 0.020 / 0 / -1.55 / 1.73 & 0.030 / 0 / -1.42 / 1.98 & \pv{0.000000e+00} \& 0.03 (N)\\
\hline

\multirow{7}{*}{User-History}
& Reputation & 717.84 / 611 / 1 / 2000 & 493.54 / 308 / 1 / 2000 & \pv{0.000000e+00} \& 0.28 (S) \\ \cline{2-5}
& UserAcceptedEditsBefore & 51.98 / 19 / 0 / 782 & 28.97 / 4 / 0 / 650 & \pv{0.000000e+00} \& 0.31 (S)\\ \cline{2-5}
& UserRejectedEditsBefore & 5.90 / 2 / 0 / 133 & 4.86 / 1 / 0 / 129 & \pv{0.000000e+00} \& 0.13 (N)\\ \cline{2-5}
& UserTotalEditsBefore & 57.88 / 22 / 0 / 782 & 33.83 / 5 / 0 / 703 & \pv{0.000000e+00} \& 0.30 (S)\\ \cline{2-5}
& UserAcceptanceRate & 0.724 / 0.833 / 0 / 0.999 & 0.501 / 0.667 / 0 / 0.997 & \pv{1.427329e-15} \& 0.38 (M)\\ \cline{2-5}
& UserRejectionRate & 0.099 / 0.073 / 0 / 0.933 & 0.126 / 0.075 / 0 / 0.938 & \pv{0.000000e+00} \& 0.01 (N)\\ \cline{2-5}
& UserEdits24h & 3.55 / 1 / 0 / 312 & 2.19 / 0 / 0 / 282 & \pv{0.000000e+00} \& 0.17 (S)\\
\hline

\end{tabular}
\end{adjustbox}
\vspace{-0.5cm}
\end{table*}


\subsection{Results}

Approximately one in six suggested edits is rejected. Among the 3,888,805 suggested edits in our dataset, 609,495 (15.67\%) are 
rejected, while 3,279,310 (84.33\%) are accepted. Although most 
suggested edits are accepted, the rejection of nearly one-sixth of 
all submissions represents a substantial portion of the review 
workload. This motivates our investigation of whether rejected edits 
exhibit characteristics that distinguish them from accepted edits. 
Table~\ref{tab:pvalues_featureset} presents the statistical comparison 
between the two groups.

\textbf{Most post-content features show small and negligible differences 
between accepted and rejected edits.} Accepted edits are associated 
with posts containing more popular tags than rejected edits (mean: 
587,750 vs.\ 445,487), with a small effect size ($|\delta| = 0.19$). 
Although post length, number of tags, number of links, and number of 
code lines also differ significantly after Bonferroni correction, 
their effect sizes are negligible ($|\delta| \leq 0.04$) and small ($|\delta| \leq 0.24$), indicating 
that post-content characteristics provide little practical separation 
between the two groups.

\textbf{Rejected edits tend to occur on posts with higher community 
visibility.} Posts receiving rejected edits have higher average scores 
(40.29 vs.\ 10.63) and more upvotes (40.94 vs.\ 11.14), both with 
small effect sizes ($|\delta| = 0.20$). This suggests that rejected 
edits are more likely to target posts that have received greater 
community attention, possibly because high-visibility posts attract 
a wider range of contributors including those less familiar with 
editing norms.

\textbf{Rejected edits tend to make larger textual changes than 
accepted edits.} Rejected edits change more body content than 
accepted edits (mean: 46.43 vs.\ 16.39 characters; $|\delta| = 
0.16$) and have a larger overall change (140.75 vs.\ 69.48 
characters; $|\delta| = 0.15$), both with small effect sizes. This 
suggests that edits involving broader textual modifications are more 
likely to be rejected, possibly because larger changes have greater 
potential to introduce unnecessary or inappropriate modifications.

\textbf{Readability changes do not meaningfully distinguish accepted 
from rejected edits.} Although all eight readability features differ 
significantly between the two groups after Bonferroni correction, all 
have negligible effect sizes ($|\delta| \leq 0.03$). For example, 
the mean Flesch readability change is 2.41 for accepted edits and 
1.13 for rejected edits, but its effect size is only $|\delta| = 
0.01$. Readability changes therefore provide little practical 
separation between the two groups.

\textbf{User editing history is the strongest signal for 
distinguishing accepted from rejected edits.} Users whose edits are 
accepted have a significantly higher prior acceptance rate than users 
whose edits are rejected (0.724 vs.\ 0.501), with the largest effect 
size among all studied features ($|\delta| = 0.38$, medium). Accepted 
edits are also associated with more prior accepted edits (51.98 
vs.\ 28.97; $|\delta| = 0.31$), more total prior edits (57.88 
vs.\ 33.83; $|\delta| = 0.30$), higher reputation (717.84 vs.\ 493.54; 
$|\delta| = 0.28$), and more suggested edits submitted within the 
previous 24 hours (3.55 vs.\ 2.19; $|\delta| = 0.17$). The mean historical acceptance and rejection rates do not
necessarily sum to one because contributors without prior
suggested edits are assigned zero for both features. These results 
confirm that contributor history provides a meaningful signal for 
differentiating accepted and rejected suggested edits, and that 
prior acceptance rate shows the largest observed effect size among the studied features.

\begin{tcolorbox}[colback=gray!8, colframe=gray!40, 
                  boxrule=0.4pt, left=6pt, right=6pt,
                  top=4pt, bottom=4pt]
\small\textbf{RQ1 Summary.} Approximately one in six suggested edits 
is rejected. All 35 features differ significantly between accepted 
and rejected edits after Bonferroni correction, but most differences 
are negligible in practical terms. User-history features show the 
strongest distinctions: prior acceptance rate has the largest effect 
size ($|\delta| = 0.38$), followed by prior accepted edits 
($|\delta| = 0.31$), total prior edits ($|\delta| = 0.30$), and 
reputation ($|\delta| = 0.28$). Rejected edits also tend to target 
posts with higher scores and to introduce larger textual changes, 
although these differences have small effect sizes.
\end{tcolorbox}

%% file: rq2_new_v2.tex
\section{\RQTwo}
\label{sec:manual-analysis}

\subsection{Motivation}
Understanding why suggested edits are rejected is important for identifying the recurring problems that make proposed edits unacceptable to reviewers. Rejection can be particularly discouraging for users, specifically for the newcomers, who may still be learning the community's editing norms and expectations. Clearer explanations of rejection reasons can help contributors understand what went wrong and better align future edits with community expectations. However, the reasons behind rejection are not well understood at a fine-grained level.

Although Stack Overflow  provides predefined rejection reasons that reviewers can select when rejecting an edit, these broad labels may encompass different types of problems and may not provide enough detail to explain the specific issue with an edit. As a result, contributors may remain uncertain about why their edits were rejected. For example, one contributor questioned a rejection by stating, \emph{``I honestly do not agree that it doesn't improve the quality of the post.''}\footnote{https://meta.stackoverflow.com/questions/433488/requesting-clarification-on-reason-for-edit-rejection} The contributor argued that the proposed formatting changes improved readability, illustrating that the selected rejection reason did not clearly communicate what reviewers considered problematic.

Such cases motivate a closer examination of rejected edits beyond their predefined rejection labels. We therefore manually analyze rejected suggested edits to identify the specific problems underlying these broad reasons and organize them into more fine-grained and interpretable categories with representative examples.

\subsection{Approach}

To obtain a more fine-grained understanding of rejection reasons, we manually examine rejected suggested edits. We determine the sample size using a 99\% confidence level and a 5\% margin of error. Based on these criteria, we randomly select 690 rejected suggested edits from our dataset for manual analysis.

\smallskip\noindent\textbf{Coding guideline construction.}
The first author (A1) initially examines a random subset of 100 sampled edits using open coding, without relying on Stack Overflow’s predefined rejection categories. For each edit, A1 compares the pre-edit and proposed post-edit versions and assigns a descriptive label representing the primary problem observed in the edit. For example, an edit that introduces a promotional URL is labeled \emph{Spam Link}, while an edit that changes valid code into syntactically invalid code is labeled \emph{Incorrect Syntax}. When the reason is unclear from the edit diff alone, A1 also examines the surrounding post context.

A1 and the second author (A2) then review the initial labels together. Semantically similar labels are merged and coding boundaries are clarified. For example, \emph{Unnecessary Reformatting} and \emph{Cosmetic Change} are consolidated into \emph{Minor Format Only Change}. The authors also establish rules for distinguishing closely related categories. This process results in an initial coding guideline containing the category name, description, representative examples, and rules for resolving ambiguous cases.

\smallskip\noindent\textbf{Independent coding.}
Using the coding guideline, A1 and A2 independently code all the 690 sampled edits. The edit diff between the pre-edit and proposed post-edit versions serves as the primary evidence. When the rejection reason cannot be determined from the diff alone, the authors additionally examine the surrounding post context. Each edit is assigned a single primary rejection reason. The coding scheme remains iterative during this process. When a recurring pattern is not adequately represented by the existing categories, A1 and A2 discuss whether to introduce a new category or refine an existing one. 

\smallskip\noindent\textbf{Agreement and disagreement resolution.}
After independently coding all 690 edits, A1 and A2 compare their assignments. Their initial agreement achieves a Cohen's kappa of $\kappa=0.80$, indicating substantial agreement. Disagreements are discussed and resolved by re-examining the edit diff, post context, and coding guidelines. When A1 and A2 cannot reach agreement, a third author (A3) reviews the case and makes the final decision. 

\subsection{Results}

\textbf{Our manual analysis identifies 17 fine-grained reasons for 
rejecting suggested edits.} From the 690 manually analyzed rejected 
edits, we identify 17 recurring rejection reasons, summarized in 
Table~\ref{tab:rejection_taxonomy}. For each reason, we provide a 
clear definition, its frequency, and representative examples. 
Detailed descriptions of each rejection reason with diff-style 
examples are presented in Appendix~A.

\textbf{Our taxonomy refines Stack Overflow 's broad rejection reasons into more 
specific and interpretable categories.}  
 Table 4 shows how our manually derived taxonomy provides finer-grained 
explanations of Stack Overflow's predefined rejection reasons. For example, 
categories such as Incorrect Syntax, Invalid Correction, and Removal 
of Essential Context explicitly identify the specific change that led 
to rejection, whereas these problems are only implicitly captured by 
broader Stack Overflow  labels such as ``invalid edit.'' Similarly, ``attempt to 
reply'' is refined into Comment on Post, Question on Post, and 
Alternate Suggestion. Spam- and vandalism-related edits are further 
distinguished as Spam Link/Keywords, Personal Promotion, Post 
Defacement, Meaningless Words, and Rude or Abusive Words. Our 
analysis also identifies recurring rejection patterns not represented 
by any dedicated Stack Overflow  rejection reason: Status Update captures edits 
that add version, deprecation, or meta-level information to a post, 
which Stack Overflow  does not provide a specific label for.

\textbf{Low-value formatting and textual changes account for more 
than half of all rejected edits.} The most frequent rejection reason 
is Minor Format Only Change, accounting for 248 edits (35.94\%). 
Undesired Text Modification accounts for another 102 edits (14.78\%), 
and Minor Grammar Only Change accounts for 13 edits (1.88\%). 
Together, these three categories represent 363 of the 690 rejected 
edits (52.61\%), indicating that many rejected suggested edits are 
not harmful but simply fail to provide sufficient improvement to 
justify modifying the post.
\input{rq2-rejections-table}
\textbf{Nearly one in five rejected edits misuses the edit mechanism 
to communicate rather than improve.} We identify 79 edits (11.45\%) 
that add a Question on Post, 38 (5.51\%) that add a Comment on Post, 
13 (1.88\%) that introduce an Alternate Suggestion, and 7 (1.01\%) 
that add a Status Update, together accounting for 137 rejected edits 
(19.86\%). These edits often contain potentially meaningful 
information, but use the suggested-edit mechanism for content that 
would be more appropriate as a comment or separate answer.

\textbf{More than one quarter of rejected edits introduce changes 
that can directly degrade post quality.} Among the 690 manually 
analyzed rejected edits, 176 fall into categories involving 
incorrect, harmful, or destructive modifications. Of these, 86 edits 
(12.46\%) alter existing content in problematic ways — such as 
modifying code incorrectly, introducing invalid corrections or 
syntax, or removing essential context. Another 90 edits (13.04\%) 
introduce inappropriate or harmful content such as spam, personal 
promotion, post defacement, meaningless text, or abusive language. 
Unlike low-value edits, these can directly affect the correctness, 
usefulness, or integrity of a post, highlighting the importance of 
identifying them for careful human review.

\begin{tcolorbox}[colback=gray!8, colframe=gray!40, 
                  boxrule=0.4pt, left=6pt, right=6pt,
                  top=4pt, bottom=4pt]
\small\textbf{RQ2 Summary.} The 17 fine-grained rejection reasons refine Stack Overflow's broad predefined rejection reasons by revealing the specific editing problems underlying them and making several implicit rejection patterns explicit. Minor formatting changes are the most common reason for rejection (35.94\%). In addition, nearly one-fifth of rejected edits (19.86\%) use the edit mechanism to add questions, comments, alternate suggestions, or status updates rather than improve the existing post. Notably, (25.51\%) of rejected edits involve potentially harmful, inappropriate, or destructive modifications, ranging from unnecessary code changes to incorrect corrections, removal of essential context, spam, and post defacement.
\end{tcolorbox}

%% file: rq2-rejections-table.tex
\begin{table}[H]
\footnotesize
\setlength{\tabcolsep}{3pt}
\renewcommand{\arraystretch}{1.15}
\centering
\caption{Taxonomy of rejected suggested edits on SO with descriptions and distribution.}
\label{tab:rejection_taxonomy}
\begin{adjustbox}{angle=90, max width=\textheight}
\begin{tabular}{|p{2cm}|p{16cm}|p{1.1cm}|}
\hline
\textbf{Rejection Reasons} & \textbf{Description} & \textbf{Count (\%)} \\
\hline

\cellcolor{gray!15}Minor Format Only Change
& \cellcolor{gray!15}Edits that alter only visual presentation without affecting meaning or functionality, such as changing indentation, adding blank lines, or switching between tabs and spaces. \textit{e.g., converting a flat code snippet into an indented block, or re-aligning import statements.}
& \cellcolor{gray!15}248 (35.94\%) \\ \hline 

Minor Grammar Only Change
& Edits that correct trivial spelling or grammatical issues that do not affect technical clarity. \textit{e.g., fixing UK vs.\ US spelling (``colour'' $\rightarrow$ ``color''), or correcting a minor typo in a title while leaving ambiguous technical content untouched.}
& 13 (1.88\%) \\ \hline 

\cellcolor{gray!15}Undesired Text Modification
& \cellcolor{gray!15}Changes to non-code textual content that do not improve clarity or correctness and may alter the author's voice or intent. \textit{e.g., replacing ``I tried X and it failed'' with ``X does not work'', or stripping a greeting such as ``Hi all,''.}
& \cellcolor{gray!15}102 (14.78\%) \\ \hline 

Undesired Code Modification
& Modifications to code that do not fix bugs or improve readability, such as renaming variables or swapping equivalent API calls without functional benefit. \textit{e.g., renaming \texttt{i} to \texttt{index} in a loop, or replacing \texttt{Array.from(x)} with \texttt{[...x]} without explanation.}
& 37 (5.36\%) \\ \hline 

\cellcolor{gray!15}Spam Link, Keywords
& \cellcolor{gray!15}Edits that insert promotional URLs, brand names, or SEO keywords unrelated to the technical content. \textit{e.g., appending ``Visit www.example.com for more tutorials'' to an answer, or embedding an affiliate link inside a code comment.}
& \cellcolor{gray!15}26 (3.77\%) \\ \hline 

Meaningless Words
& Additions of nonsensical, garbled, or irrelevant text that reduce post quality, often from copy-paste errors or automated edits. \textit{e.g., inserting ``asdfgh lorem blah'' mid-paragraph, or appending a random string of characters at the end of a working answer.}
& 37 (5.36\%) \\ \hline 

\cellcolor{gray!15}Personal Promotion
& \cellcolor{gray!15}Inclusion of self-promotional content unrelated to the post's technical substance, such as references to the editor's own library, blog, or service. \textit{e.g., adding ``I built a library that handles this---check it out at github.com/myrepo'' without the tool being directly relevant.}
& \cellcolor{gray!15}11 (1.59\%) \\ \hline 

Post Defacement
& Intentional edits that corrupt or vandalize a post by deleting substantial content or replacing it with garbage text. \textit{e.g., blanking out an entire accepted answer and replacing it with ``this is wrong'', or deleting all code from a question.}
& 11 (1.59\%) \\ \hline 

\cellcolor{gray!15}Rude or Abusive Words
& \cellcolor{gray!15}Insertion of offensive, profane, or personally insulting language targeting other users or the asker. \textit{e.g., adding ``OP is an idiot for not knowing this'' to an answer, or inserting a slur into a neutral technical explanation.}
& \cellcolor{gray!15}5 (0.72\%) \\ \hline 


Comment on Post
& Edits that insert remarks, feedback, or conversational responses directed at the original author rather than improving the post content, such as asking the author to clarify an example or commenting on the proposed solution \textit{e.g., editing a question to add ``Have you tried using \texttt{async/await}? That solved it for me.''}
& 38 (5.51\%) \\ \hline 

\cellcolor{gray!15}Status Update
& \cellcolor{gray!15}Adding contextual update notes---such as version deprecations or newer API availability---that do not directly improve the original content and are better suited to a comment or new answer. \textit{e.g., prepending ``Note: As of the date, this method is deprecated in favor of X'' without elaboration.}
& \cellcolor{gray!15}7 (1.01\%) \\ \hline 

Invalid Correction
& Edits that attempt to fix a technical error but introduce incorrect or misleading information, making the post less accurate than before. \textit{e.g., changing a correct SQL JOIN clause to an incorrect one, or replacing a valid regex pattern with a broken one.}
& 13 (1.88\%) \\ \hline 

\cellcolor{gray!15}Incorrect Syntax
& \cellcolor{gray!15}Edits that introduce syntactic errors or invalid code constructs into an otherwise correct snippet. \textit{e.g., adding a stray semicolon inside a Python function definition, or removing a closing bracket that causes a JavaScript block to fail parsing.}
& \cellcolor{gray!15}12 (1.74\%) \\ \hline 

Alternate Suggestion
& Adding an entirely different solution or approach to the post body rather than improving it, which belongs in a separate answer. \textit{e.g., appending ``Alternatively, you could use \texttt{pandas.merge()}'' to an existing SQL JOIN-based answer without the original author's consent.}
& 13 (1.88\%) \\ \hline 

\cellcolor{gray!15}Removal of essential context
& \cellcolor{gray!15}Edits that strip content essential to the post's intent or usefulness, such as deleting clarifying examples, error messages, or context that helps readers understand the problem. \textit{e.g., removing a minimal reproducible code example from a question, or deleting an explanatory paragraph that justified a design decision.}
& \cellcolor{gray!15}24 (3.48\%) \\ \hline 

Question on Post
& Adding follow-up or clarifying questions into the post body rather than using comments, turning a standalone post into an incomplete dialogue. \textit{e.g., inserting ``But what if the array is empty?'' or ``Did you also try this on Windows?'' at the end of an existing question.}
& 79 (11.45\%) \\ \hline 

\cellcolor{gray!15}Other
& \cellcolor{gray!15}Edits that do not clearly fall into any defined category or lack sufficient context for reliable classification. \textit{e.g., an edit that partially reformats code while also partially rewords a sentence, with no single dominant rejection reason.}
& \cellcolor{gray!15}14 (2.03\%) \\  \hline 
\end{tabular}
\end{adjustbox}

\end{table}

\begin{table*}[t]
\centering
\caption{How our taxonomy refines and extends SO's predefined rejection reasons.}
\label{tab:so_vs_taxonomy}
\scriptsize
\setlength{\tabcolsep}{3pt}
\renewcommand{\arraystretch}{1.08}

\begin{adjustbox}{max width=\textwidth}
\begin{tabular}{p{2.5cm}p{4.5cm}p{6.0cm}}
\hline
\textbf{SO Predefined Reason} &
\textbf{Fine-grained Reasons in Our Taxonomy} &
\textbf{Refinement / Additional Explanation} \\
\hline

Too minor / No improvement whatsoever &
\emph{Minor Format Only Change};
\emph{Minor Grammar Only Change};
\emph{Undesired Text Modification} &
Distinguishes formatting-only changes, minor grammar corrections, and unnecessary text modifications. \\

Invalid edit &
\emph{Invalid Correction};
\emph{Incorrect Syntax};
\emph{Undesired Code Modification};
\emph{Undesired Text Modification} &
Clarifies whether an edit introduces incorrect information, broken syntax, unnecessary code changes, or unnecessary text changes. \\

Attempt to reply &
\emph{Comment on Post};
\emph{Question on Post};
\emph{Alternate Suggestion} &
Separates comments, follow-up questions, and alternative solutions submitted through the edit mechanism. \\

Spam or vandalism / Vandalism &
\emph{Spam Link/Keywords};
\emph{Personal Promotion};
\emph{Post Defacement};
\emph{Meaningless Words};
\emph{Rude or Abusive Words} &
Separates promotional, destructive, meaningless, and abusive edits. \\

Radical change / Conflicts with author's intent &
\emph{Removal of Essential Context};
\emph{Undesired Text Modification} &
Distinguishes removal of important information from unnecessary changes to the author's content. \\

No dedicated SO reason &
\emph{Status Update} &
Captures version, deprecation, temporary, or meta-level information added through an edit. \\

\hline
\end{tabular}
\end{adjustbox}
\end{table*}

%% file: rq3.tex
\section{\RQThree}
\subsection{Motivation}
\label{rq3}

Our dataset shows that approximately 84.33\% of suggested edits are ultimately accepted, while only about 15.67\% are rejected. Nevertheless, all suggested edits currently require human review before they can be accepted. As a result, reviewers spend substantial effort examining a large number of edits for which they ultimately find no reason for rejection. This creates a substantial review workload, as most reviewer effort is spent on edits that are eventually accepted. Reducing manual review of low-risk edits could allow reviewer effort to be focused more effectively on problematic edits (i.e., edits that are more likely to get rejected).

However, the challenge is to reduce this workload without overlooking too many problematic edits. An effective review mechanism should therefore minimize the number of edits requiring manual inspection while still ensuring that most edits that would otherwise be rejected are brought to reviewers' attention. We therefore investigate whether the suggested-edit review mechanism can be made more efficient by reducing manual intervention while maintaining strong coverage of problematic edits.

\subsection{Approach}

To examine whether manual review workload can be reduced while still retaining problematic edits for human inspection, we consider a threshold-based setting based on the predicted rejection risk of each suggested edit~\cite{halfaker2019ores, west2010stiki, makhijani2021quest}. Edits with higher predicted rejection risk are prioritized for immediate human review, while lower-risk edits remain in the standard review queue. The predictions are therefore used to allocate reviewer attention rather than replace human moderation decisions. Under this setting, the objective is to reduce the number of edits requiring immediate expert review while still capturing a large proportion of edits that would otherwise be rejected.

\subsubsection{Evaluation techniques}
We evaluate three different techniques for estimating the rejection risk of a suggested edit: (1) a framework derived from prior work on edit management, (2) an embedding-based model that captures the semantic changes introduced by an edit and (3) a feature-based model using the characteristics studied in RQ1. We compare these techniques to determine which one best supports this review setting by reducing manual review workload while retaining problematic edits for human inspection. We briefly describe the three evaluated techniques below.

\begin{enumerate}

\item \textbf{EditEx Model.}
The first technique is based on the EditEx framework proposed by Mondal et al.~\cite{autopredsaikat1}. Their study formulated rollback prediction as a binary classification problem using structural and contextual edit features. They evaluated Decision Tree (DT), Random Forest (RF), k-Nearest Neighbors (KNN), Support Vector Machine (SVM), and Extreme Gradient Boosting (XGBoost), with Random Forest achieving the best performance on their dataset.

We adopt the EditEx feature design for suggested-edit rejection prediction. Rather than using only the predicted class label, we use the model's rejection probability for each edit, denoted as \(P_i^{(\mathrm{EditEx})}\). A higher probability indicates that the edit is more likely to be rejected. These probabilities are subsequently used by the threshold-based review policy described in Section~\ref{sec:review_policy}.

\item \textbf{Embedding Model.}
The second technique represents each suggested edit using a dense embedding of the changes introduced by the edit. We include this technique because embedding-based representations can capture semantic information and have been successfully applied to software-engineering tasks involving documents and code changes~\cite{codechange,llm2vec,llm4code}.

For each suggested edit, we extract the difference between the pre-edit and proposed post-edit versions using Python's \texttt{difflib.ndiff}\footnote{\url{https://docs.python.org/3/library/difflib.html}}. We encode the resulting diff using \texttt{gte-Qwen2-7B-instruct} through the Hugging Face Transformers library~\cite{alibabanlp2024gteqwen2,qwen2}. We obtain a 3,584-dimensional representation by mean-pooling the final hidden-layer token representations. The embedding model is used without fine-tuning.

We then train a Multi-Layer Perceptron (MLP) classifier on these embeddings using \texttt{scikit-learn}~\cite{scikit-learn}. The MLP contains one hidden layer with 256 ReLU units and is optimized using Adam with a learning rate of \(10^{-4}\) and a batch size of 256. Training runs for at most 100 epochs, with early stopping after 10 consecutive epochs without improvement in validation loss. The model outputs a rejection probability \(P_i^{(\mathrm{Embedding})}\) for each suggested edit. All embedding experiments are conducted using an NVIDIA Tesla K80 GPU.

\item \textbf{Feature Model.}
The third technique uses the 35 structured features introduced in RQ1. These features characterize five dimensions of a suggested edit: post content, post history, change magnitude, readability variation, and user history. The features are extracted using the procedure described in Section~\ref{sec:feature_collection}.

We normalize numeric features using \texttt{MinMaxScaler} from \texttt{scikit-learn}~\cite{scikit-learn}. We evaluate several classification algorithms, including DT, RF, KNN, SVM, XGBoost, and MLP, implemented using \texttt{scikit-learn} and the \texttt{XGBoost} library~\cite{scikit-learn,xgboost}. Based on our cross-validation comparison, XGBoost achieves the best performance among the evaluated classifiers and is therefore used as the Feature Model.

XGBoost is trained using binary logistic loss and produces a rejection probability \(P_i^{(\mathrm{Feature})}\) for each edit. As with the other techniques, we use this probability in the threshold-based review policy rather than directly using the predicted class label.

\end{enumerate}

\subsubsection{Threshold-based Review Policy}
\label{sec:review_policy}

We use the predicted rejection probabilities to determine which suggested edits require manual review. Given a threshold $\tau$, an edit with a predicted rejection probability greater than or equal to $\tau$ is sent to human reviewers. High-risk edits are prioritized for immediate human review, whereas lower-risk edits (i.e., an edit with a probability below $\tau$) can remain in the regular review queue and, depending on the platform’s risk tolerance, may be automatically accepted after a configurable waiting period if no reviewer intervenes. Thus, the threshold controls the proportion of submitted edits that require reviewer intervention.


The threshold determines the trade-off between reviewer workload and coverage of problematic edits. A lower threshold sends more edits to reviewers, increasing the likelihood that edits that would otherwise be rejected receive human inspection, but reducing the potential workload savings. A higher threshold sends fewer edits to reviewers and therefore reduces manual review workload, but increases the proportion of rejected edits that are not prioritized for immediate review. We evaluate the three techniques across different thresholds to determine how effectively they balance these two objectives.

\subsubsection{Evaluation Procedure}
Our collected suggested edits data covers the period from 2011 to May 2023. To build and evaluate our predictive models, we followed prior study and constructed a time-aware split between training and testing data \cite{autopredsaikat1}. From the edits submitted between 2011 and 2022, we randomly sampled 100,000 edits for training, drawing approximately the same number of edits from each year. For testing, we used 20,236 edits from the most recent five-month period in our dataset, covering January 2023 to May 2023. This temporal separation ensures that older edits are used to train the model, while newer edits are used to evaluate its performance.

\begin{figure*}[t]
\centering
\noindent

\begin{minipage}[t]{1.0\textwidth}
\centering
\vspace{0pt}

\captionsetup{type=table, width=\linewidth}
\caption{Review rate and recall comparison across all techniques at different decision thresholds. Highlighted rows ($\tau$ = 0.40–0.60) indicate practical operating points that balance the proportion of edits prioritized for human review with the recall of rejected edits.}
\label{tab:aggressive_all}
\vspace{-2pt}
\footnotesize
\setlength{\tabcolsep}{4pt}
\renewcommand{\arraystretch}{1.0}
\begin{tabular}{|c|cc|cc|cc|}
\hline 
& \multicolumn{2}{c|}{\textbf{Feature Model}} 
& \multicolumn{2}{c|}{\textbf{Embedding Model}} 
& \multicolumn{2}{c|}{\textbf{EditEx Model \cite{autopredsaikat1}}} \\
\cline{2-7}
Decision Threshold($\tau$) & Review Rate & Recall & Review Rate & Recall & Review Rate & Recall \\
\hline
0.10 & 96.4\% & 99.3\% & 100.0\% & 100.0\% & 94.4\% & 96.9\% \\
0.20 & 83.5\% & 95.5\% & 99.9\%  & 100.0\% & 84.9\% & 91.1\% \\
0.30 & 67.1\% & 89.1\% & 90.9\%  & 95.7\%  & 72.3\% & 82.1\% \\
0.40 & \hl{\textbf{52.3\%}} & \hl{\textbf{80.8\%}} & \textbf{60.9\%}  & \textbf{78.8\%}  & \textbf{59.0\%} & \textbf{72.5\%} \\
0.50 & \hl{\textbf{39.6\%}} & \hl{\textbf{71.8\%}} & \textbf{36.8\%}  & \textbf{59.1\%}  & \textbf{46.3\%} & \textbf{62.1\%} \\
0.60 & \hl{\textbf{30.3\%}} & \hl{\textbf{63.2\%}}& \textbf{21.8\%} & \textbf{40.6\%} & \textbf{33.9\%} & \textbf{50.1\%} \\
0.70 & 22.6\% & 53.5\% & 10.4\%  & 23.0\%  & 23.3\% & 40.7\% \\
0.80 & 14.6\% & 41.7\% & 3.1\%   & 8.7\%   & 14.5\% & 29.6\% \\
0.90 & 6.7\%  & 23.1\% & 0.3\%   & 0.9\%   & 7.5\%  & 18.4\% \\
\hline
\end{tabular}
\label{rq3_results}
\end{minipage}%
\hfill
\end{figure*}

\subsubsection{Evaluation Metric}

To evaluate how well the model prioritizes rejected edits, we apply a 
decision threshold $\tau$ to the predicted rejection scores. Edits with scores greater than or equal to $\tau$ are flagged as likely to be rejected and prioritized for immediate human inspection to maintain content quality. Edits with scores below $\tau$ remain in the standard review queue but receive lower review priority. We measure performance using two metrics: prioritized review rate and recall. Thus, a lower review rate indicates that a smaller proportion of submissions requires prioritized human attention.

The \textbf{review rate} measures the proportion of all submitted edits that the model flags as likely to be rejected and therefore routes to human reviewers for further inspection. For example, if the model flags 300 out of 1,000 submitted edits, the review rate is 30\%, meaning that reviewers must inspect 30\% of all submissions. Thus, a lower review rate indicates a smaller human-review workload, whereas a higher review rate requires reviewers to inspect a larger proportion of the submitted edits.

\begin{equation}
\label{eq:review_rate}
\text{review rate} =
\frac{\text{Number of edits flagged for human review}}
{\text{Total number of submitted edits}}
\end{equation}



The \textbf{recall} measures the proportion of the truly rejected edits 
the model successfully captures. If there are 200 truly rejected edits and 
the model identifies 140 of them as likely to be rejected at a given threshold, the recall is 70\%. A higher recall means 
fewer rejected edits slip through undetected.

\begin{equation}
\label{eq:recall}
\text{Recall} =
\frac{\text{Number of edits identified as likely to be rejected by the model}}
     {\text{\ Total number of rejected edits}}
\end{equation}

Together, \emph{review rate} and \emph{recall} describe the trade-off between reviewer 
effort and rejection detection. A good model achieves high \emph{recall} while 
keeping the \emph{review rate low}, meaning it captures a large share of 
rejected edits without requiring reviewers to manually inspect too many 
submissions. For example, a model that achieves 70\% \emph{recall} at a 30\% 
\emph{review rate} captures 70\% of all rejected edits while only flagging 30\% 
of total submissions for immediate review.

\subsection{Results}

\textbf{The Feature Model provides the best balance between recall 
and review rate at moderate thresholds.} 
Table~\ref{tab:aggressive_all} compares the three techniques across 
different decision thresholds. At $\tau = 0.30$, the Feature Model 
captures 89.1\% of rejected edits while sending only 67.1\% of all 
edits for human review. The Embedding Model achieves higher recall 
(95.7\%) but requires a substantially higher review rate of 
90.9\%, while the EditEx Model achieves 82.1\% recall with a human 
review rate of 72.3\%. At $\tau = 0.40$, the Feature Model achieves 
80.8\% recall with a review rate of 52.3\%, compared with 
78.8\% recall and 60.9\% review rate for the Embedding Model, 
and 72.5\% recall and 59.0\% review rate for the EditEx Model. 
This means the Feature Model captures more rejected edits while 
requiring human review for fewer submissions than both baselines.

\textbf{Very low thresholds capture almost all rejected edits but 
offer little reduction in review effort.} At $\tau = 0.10$, the 
Embedding and Feature Models achieve recalls of 100.0\% and 99.3\% 
respectively, but also send 100.0\% and 96.4\% of all edits for 
human review. A similar pattern occurs at $\tau = 0.20$, where all 
three techniques still require human review for more than 80\% of 
submissions. Although low thresholds provide high recall, they offer 
little practical benefit for reducing reviewer workload.

\textbf{The advantage of the Feature Model becomes more pronounced 
as fewer edits are prioritized for review.} At $\tau = 0.50$, the 
Feature Model requires human review for only 39.6\% of submitted 
edits while capturing 71.8\% of rejected edits. The Embedding Model 
requires a similar review rate (36.8\%) but captures only 
59.1\% of rejected edits, while the EditEx Model requires 46.3\% 
review rate and captures 62.1\%. At $\tau = 0.60$, the Feature 
Model further reduces the review rate to 30.3\% while 
maintaining 63.2\% recall, compared to 40.6\% and 50.1\% recall for 
the Embedding and EditEx Models respectively. These results show that 
rejected edits can be effectively concentrated near the top of the 
review queue, allowing reviewers to focus first on a smaller subset 
of potentially problematic submissions.

\textbf{The Feature Model retains more rejected edits even under 
strict review thresholds.} As $\tau$ increases beyond 0.60, recall 
decreases for all three techniques. However, the Feature Model 
consistently retains higher recall than the other two studied models. At $\tau = 
0.80$, it captures 41.7\% of rejected edits while requiring human 
review for only 14.6\% of submissions. At a comparable human review 
rate, EditEx captures only 29.6\% and the Embedding Model captures 
only 8.7\%, showing that the Feature Model concentrates rejected 
edits more effectively even under aggressive filtering.

\subsection{Comparison with State-of-the-Art LLM Baselines}
After evaluating the three primary techniques on the main held-out test set, we conduct an additional experiment to examine how the best-performing Feature Model compares with general-purpose LLMs under zero-shot classification. Due to the financial cost of querying LLMs at scale, we evaluate all models 
on a stratified random sample of 500 edits drawn from the test set, balanced 
equally between accepted (250) and rejected (250) edits. We note that this 
balanced setting differs from the natural class distribution used in RQ3 
(approximately 15.67\% rejected, 84.33\% accepted). Consequently, results reported 
here are not directly comparable to Table~\ref{rq3_results} and should 
be interpreted as a relative comparison among models on the same sample 
rather than an absolute performance estimate under realistic deployment 
conditions. To ensure a fair comparison, we re-evaluate our feature-based 
XGBoost model on the same 500 balanced samples under identical conditions.

Recent software engineering research has increasingly adopted large language 
models as zero-shot baselines for classification and prioritization 
tasks~\cite{yang2024large, li2024codereviewer, sun2024llm4vuln, 
niu2024empirical}. To situate our feature-based approach within this 
landscape, we conduct a complementary evaluation comparing our model against 
three state-of-the-art LLMs under a zero-shot setting: 
GPT-4o~\cite{openai2024gpt4}, Qwen2.5-32B~\cite{qwen2024}, and 
Llama-3.3-70B~\cite{meta2024llama33}. These models represent three dominant 
families currently used as baselines in SE research: proprietary 
general-purpose models with strong reasoning capabilities 
(GPT-4o)~\cite{openai2024gpt4, white2023chatgpt}, open-source large-scale 
models with competitive benchmark performance 
(Qwen2.5-32B)~\cite{qwen2024}, and open-source 
instruction-tuned models widely adopted in empirical SE studies 
(Llama-3.3-70B)~\cite{dubey2024llama, niu2024empirical}.

\subsubsection*{Prompt Design} 
\begin{figure}[ht]
    \centering
    \includegraphics[width=\textwidth]{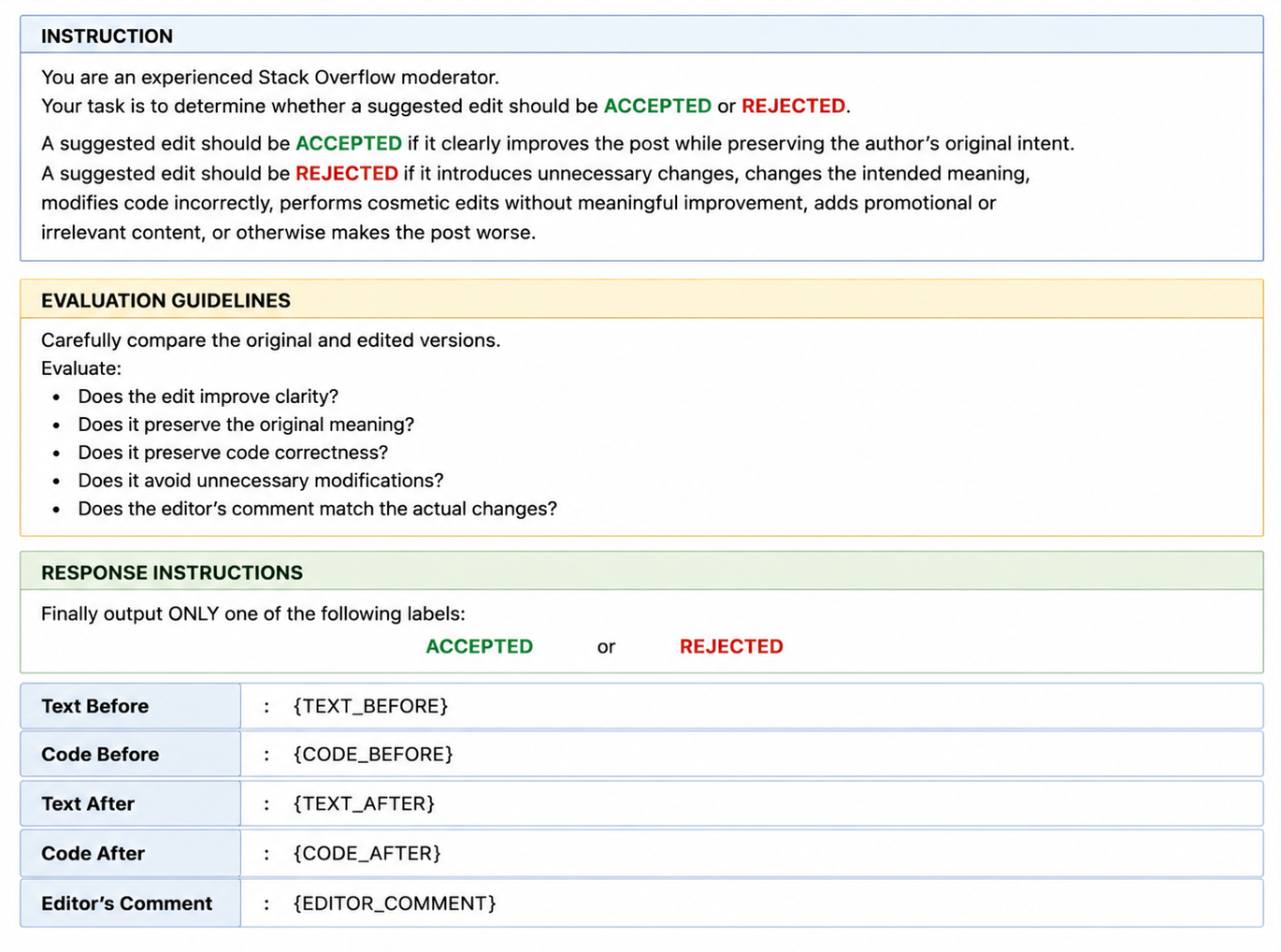}
    \caption{Zero-shot prompt template used for suggested edit classification. The prompt provides the original and edited text, code, and the editor's comment, and instructs the LLM to predict whether the suggested edit should be ACCEPTED or REJECTED.}
    \label{fig:prompt}
\end{figure}

For each sampled edit, we construct a zero-shot prompt containing five 
components: the original post text before the edit, the original code before 
the edit, the post text after the edit, the code after the edit, and the 
editor's comment describing the intent of the edit. The full prompt template 
is shown in Figure~\ref{fig:prompt}. Each model is asked to predict whether 
the suggested edit will be accepted or rejected by Stack Overflow  reviewers, 
producing a binary classification decision. For our feature-based model, we 
additionally calibrate predictions across probability thresholds $\tau$ = 
0.4, 0.5, and 0.6 to identify the operating point that maximizes recall on 
the rejected class, since our primary goal is to capture as many 
review-rejected edits as possible to keep the knowledge source clean and 
reliable.

\subsubsection*{Results}
Table~\ref{tab:llm_comparison} presents classification performance for all 
four models on the balanced 500-edit sample. Our feature-based model achieves 
the highest accuracy (64.6\%), the highest rejected-class recall (67.6\%), 
and the highest rejected-class F1-score (0.656), outperforming all three LLM 
baselines on metrics (e.g., recall, F1-score) relevant to our moderation goal. Among the LLM 
baselines, Qwen2.5-32B achieves the highest overall accuracy (57.2\%) and 
the highest rejected-class precision (71.9\%), but at the cost of very low 
rejected-class recall (23.6\%), meaning it misses the large majority of 
rejected edits. GPT-4o and Llama-3.3-70B achieve similar rejected-class 
recall (31.2\% and 29.6\% respectively), both substantially below our 
feature-based model.

\begin{table}[t]
\centering
\caption{Classification performance comparison on the balanced 500-edit 
sample (250 accepted, 250 rejected). The best value in each column is shown 
in \textbf{bold}.}
\label{tab:llm_comparison}
\footnotesize
\setlength{\tabcolsep}{4pt}
\begin{tabular}{llcccc}
\toprule
\textbf{Model} & \textbf{Class} & \textbf{Precision} & \textbf{Recall} 
& \textbf{F1} & \textbf{Accuracy} \\
\midrule
\multirow{2}{*}{GPT-4o}
  & Accepted & 0.526 & 0.764 & 0.623 & \multirow{2}{*}{0.538} \\
  & Rejected & 0.569 & 0.312 & 0.403 & \\
\midrule
\multirow{2}{*}{Qwen2.5-32B}
  & Accepted & 0.543 & 0.908 & 0.680 & \multirow{2}{*}{0.572} \\
  & Rejected & 0.720 & 0.236 & 0.355 & \\
\midrule
\multirow{2}{*}{Llama-3.3-70B}
  & Accepted & 0.532 & 0.800 & 0.639 & \multirow{2}{*}{0.548} \\
  & Rejected & 0.597 & 0.296 & 0.396 & \\
\midrule
\multirow{2}{*}{\textbf{Feature Model ($\tau$=0.5)}}
  & Accepted & \textbf{0.655} & 0.616 & 0.635 
  & \multirow{2}{*}{\textbf{0.646}} \\
  & Rejected & 0.638 & \textbf{0.676} & \textbf{0.656} & \\
\bottomrule
\end{tabular}
\end{table}

The confusion matrices further illustrate the gap. All three LLM baselines classify the large majority of rejected edits as accepted: GPT-4o misses 172 out of 250 rejected edits, Llama-3.3-70B misses 176, and Qwen2.5-32B misses 191. In contrast, our feature-based model at $\tau$ = 0.5 misses only 81 rejected edits out of 250, correctly identifying 169. This difference is critical in the context of our moderation goal: an edit that slips through undetected may degrade the quality of Stack Overflow's knowledge base and increases downstream reviewer burden.

\subsubsection*{Threshold Calibration}
Since our primary objective is to maximize detection of rejected edits, we additionally calibrate our feature-based model across three probability thresholds and report rejected-class performance in Table~\ref{tab:threshold_calibration}. Lowering the threshold from 0.5 to 0.4 increases rejected-class recall from 67.6\% to 78.0\%, at the cost of reduced precision (60.2\%) and a slight drop in overall accuracy (63.2\%). Raising the threshold to 0.6 improves precision to 69.5\% but reduces recall to 58.4\%. For deployment scenarios where missing a rejected edit is more costly than flagging an accepted edit for additional review, $\tau$ = 0.4 represents the optimal operating point, capturing 195 out of 250 rejected edits while maintaining acceptable precision.

\begin{table}[t]
\centering
\caption{Threshold calibration for the feature-based model on the balanced 
500-edit sample.}
\label{tab:threshold_calibration}
\footnotesize
\setlength{\tabcolsep}{5pt}
\begin{tabular}{ccccc}
\toprule
$\tau$ & \textbf{Accuracy} & \textbf{Rejected Precision} & \textbf{Rejected Recall} 
& \textbf{Rejected F1} \\
\midrule
0.4 & 0.632 & 0.602 & \textbf{0.780} & 0.679 \\
0.5 & 0.646 & 0.638 & 0.676 & 0.656 \\
0.6 & \textbf{0.664} & \textbf{0.695} & 0.584 & 0.635 \\
\bottomrule
\end{tabular}
\end{table}

The consistent underperformance of all three LLM baselines on the rejected class can be attributed to a limitation of the zero-shot setting: LLMs operate on textual and code content alone and have no access to the user-history signals that our RQ1 analysis identified as the strongest discriminators between accepted and rejected edits — specifically prior acceptance rate, prior rejection rate, and total submission history. Without these behavioral signals, LLMs must infer rejection likelihood from edit content alone, which our feature analysis confirms is a comparatively weak predictor of reviewer decisions. In our evaluation, structured behavioral features outperform the zero-shot LLM baselines on rejected-edit detection. Among the LLM baselines, Qwen2.5-32B shows a strong tendency to classify edits as accepted (predicting acceptance for 418 out of 500 edits), suggesting a conservative bias toward approval that is misaligned with the moderation goal. GPT-4o and Llama-3.3-70B show a similar but slightly less extreme pattern. These observations suggest that fine-tuning LLMs on suggested-edit data with explicit user history context, or combining LLM semantic scoring with our feature-based behavioral signals in a hybrid architecture, could yield further 
improvements and represents a promising direction for future work.

\begin{tcolorbox}[colback=gray!8, colframe=gray!40,
                  boxrule=0.4pt, left=6pt, right=6pt,
                  top=4pt, bottom=4pt]
\small\textbf{RQ3 Summary.} The Feature Model provides the strongest 
trade-off between review rate and rejected-edit recall among the three 
evaluated techniques. At $\tau = 0.40$, it prioritizes 52.3\% of edits 
for human review while capturing 80.8\% of rejected edits, compared 
with 60.9\% review rate and 78.8\% recall for the Embedding Model and 
59.0\% review rate and 72.5\% recall for the EditEx Model. Even at 
$\tau = 0.80$, the Feature Model prioritizes only 14.6\% of edits while 
capturing 41.7\% of rejected edits. In a complementary zero-shot 
evaluation on another 500 edits, the Feature Model achieves 78.0\% rejected-edit 
recall at $\tau = 0.40$, compared with 23.6--31.2\% for GPT-4o, 
Qwen2.5-32B, and Llama-3.3-70B. Overall, the results show that structured 
edit, post, and contributor-history features can effectively prioritize 
potentially problematic suggested edits for human review.
\end{tcolorbox}

%% file: 044-Implications.tex
\section{Discussion}
\label{discussion}
Our findings reveal several important implications for Stack Overflow  
stakeholders, spanning platform design, reviewer practice, contributor 
experience, and future research directions.

\subsection{Implications for Platform Designers}

Our analysis identifies two structural issues that contribute to reviewer workload and contributor frustration independently of edit quality. First, while examining Stack Overflow's rejection reasons, we found that \emph{suggested-edit conflicts}---procedural rejections caused by concurrent editing---account for approximately 20\% of all rejected edits in our dataset. As 
reported in Section~\ref{sec:data_collection}, these rejections are not content-quality 
judgments; they are unavoidable race conditions under the current system 
design. Contributors, especially newcomers, may invest significant effort 
improving a post only to receive an automated rejection for reasons entirely 
unrelated to the quality of their work~\footnote{https://meta.stackoverflow.com/questions/401701/edits-that-conflict-with-a-subsequent-edit-should-not-count-as-rejected-edits}. 
Platform designers should consider notifying contributors earlier when 
concurrent edits are detected, or providing mechanisms to reapply a pending 
suggested edit to the updated version of a post, thereby preserving 
contributor effort and reducing demotivation.

Second, the predefined rejection labels available to reviewers are broad and 
sometimes ambiguous, as demonstrated in Section~\ref{sec:manual-analysis}. Labels such as 
``no improvement'' or ``too minor'' cover a wide range of situations and 
provide little actionable guidance to contributors about how to improve future 
edits. Our refined taxonomy of 17 fine-grained rejection reasons, each with 
concrete examples, provides a foundation for improving the reviewer interface. 
Incorporating structured rejection reasons with illustrative examples directly 
into the review queue could reduce reviewer inconsistency, improve contributor 
understanding, and increase the perceived fairness of moderation decisions.

\subsection{Implications for Reviewers}

Our findings show that formatting-related edits occupy a persistent gray area in reviewer judgment. A large proportion of both accepted and rejected edits involve formatting adjustments or minor grammar corrections, and reviewers frequently reach inconsistent decisions on edits that are structurally similar. This inconsistency increases cognitive load for reviewers and creates confusion for contributors who receive contradictory feedback across submissions. Reviewers should be aware that their decisions on borderline formatting edits carry disproportionate consequences for newcomers who rely on this feedback to learn community norms. Clearer reviewer guidelines specifically targeting formatting, for example, distinguishing between formatting changes that improve code readability and those that introduce purely cosmetic variation, would reduce inconsistency and improve the reliability of the review process.

Additionally, our feature analysis shows that the largest differences between accepted and rejected edits occur in contributor-history characteristics, particularly prior acceptance rate, number of prior accepted edits, total submission history, and reputation. Reviewers who consciously or unconsciously weigh reputation heavily in their decisions may be introducing bias against low-reputation contributors whose individual edits are substantively sound. Our results suggest that content-based signals and contributor editing history are more reliable indicators of edit quality than raw reputation score.

\subsection{Implications for Contributors}
\begin{figure}[t]
\centering
\begin{tikzpicture}
\begin{axis}[
    xlabel={User Reputation Group},
    ylabel={Percentage (\%)},
    ybar,
    bar width=10pt,
    width=\columnwidth,
    height=6cm,
    symbolic x coords={
        0--200,201--400,401--600,601--800,801--1000,
        1001--1200,1201--1400,1401--1600,1601--1800,1801--2000
    },
    xtick=data,
    xticklabel style={
        rotate=45,
        anchor=east,
        font=\scriptsize
    },
    ymin=0,
    ymax=110,
    enlarge x limits=0.05,
    legend style={
        at={(0.5,1.12)},
        anchor=north,
        legend columns=2,
        font=\small
    },
    ymajorgrids=true,
    grid style=dashed,
]
\addplot[
    fill=gray!30,
    draw=black
] coordinates {
    (0--200,24.20)
    (201--400,12.00)
    (401--600,13.40)
    (601--800,10.00)
    (801--1000,10.20)
    (1001--1200,8.60)
    (1201--1400,6.60)
    (1401--1600,6.20)
    (1601--1800,6.00)
    (1801--2000,2.80)
};
\addplot[
    fill=black!75,
    draw=black
] coordinates {
    (0--200,66.12)
    (201--400,78.33)
    (401--600,86.57)
    (601--800,88.00)
    (801--1000,80.39)
    (1001--1200,88.37)
    (1201--1400,93.94)
    (1401--1600,88.89)
    (1601--1800,90.00)
    (1801--2000,100.00)
};
\legend{Edit Submission Share (\%), Acceptance Rate (\%)}
\end{axis}
\end{tikzpicture}
\caption{Distribution of suggested-edit submissions and acceptance rates 
across user reputation groups. The submission share shows that a large 
proportion of edits (24.20\%) come from the lowest reputation group 
(0--200), with submission volume declining steadily as reputation increases. 
Acceptance rate generally trends upward with reputation but is non-monotonic, 
peaking at 93.94\% for the 1201--1400 group before dipping slightly in the 
1401--1800 range and reaching 100\% only in the smallest, highest-reputation 
group (1801--2000).}
\label{fig:rep_stats}
\end{figure}
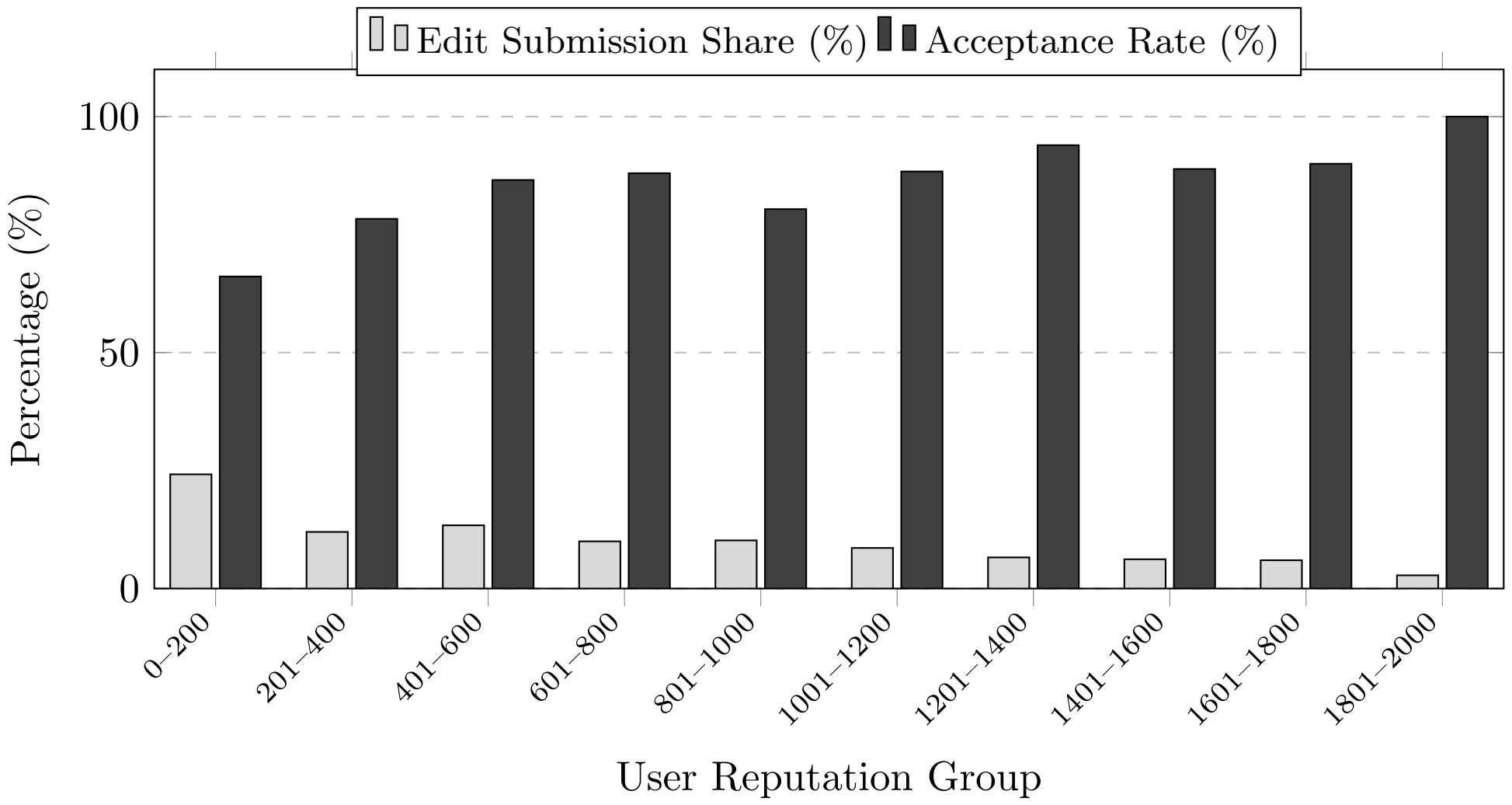

Figure~\ref{fig:rep_stats} shows the relationship between user reputation and 
acceptance rate for suggested edits. Mondal et al.~\cite{autopredsaikat1} found 
that for rollback edits, reputation is the most important feature in their 
random forest model — community trust, as estimated by reputation score, is a 
strong predictor of edit acceptability, with high-reputation users consistently 
more likely to have their edits accepted. Our analysis of suggested edits 
reveals a more nuanced pattern. While acceptance rate is lowest for the 
lowest-reputation group (66.12\% for users with 0--200 reputation) and 
generally trends upward with reputation, the relationship is not uniform: 
acceptance rate peaks at 93.94\% for the 1201--1400 reputation group before 
declining to 88.89\% for the 1401--1600 group and continuing to fluctuate in 
the 1601--1800 range, prior to reaching 100\% in the small top group 
(1801--2000). This non-monotonic pattern suggests that reputation alone does 
not fully explain acceptance outcomes for suggested edits in the way it does 
for rollback edits, and that other factors — such as edit scope and content 
quality — interact with reputation in ways not captured by a simple linear 
trust relationship. Contributors with low reputation should be 
aware that edit scope and specificity matter: our RQ1 findings show that edits 
involving large or disruptive text changes are more likely to be rejected 
regardless of intent, while targeted improvements with clear and localized 
changes are more consistently accepted. Building a positive editing history — 
evidenced by our finding that prior acceptance rate shows the strongest association with review outcome among the studied features — remains the most reliable long-term strategy 
for improving review outcomes, particularly since reputation alone does not 
guarantee acceptance.

\subsection{Implications for Researchers}

Our study demonstrates that probability-based prioritization of the suggested-edit 
review queue is both feasible and effective under realistic class imbalance. 
However, several open problems remain. First, the overlap between accepted and 
rejected edits in the feature space — particularly for formatting-related edits 
— suggests that structural features alone are insufficient to resolve borderline 
cases. Future work should investigate fine-tuned language models trained 
specifically on suggested-edit content as a complement to feature-based 
approaches, potentially providing more reliable quality signals for ambiguous 
edits. Second, our taxonomy of 17 rejection reasons provides a labeled 
framework that could support automated rejection reason prediction, moving 
beyond binary outcome classification toward more interpretable and actionable 
moderation assistance. Third, the reputation bias observed in acceptance 
patterns warrants dedicated investigation: disentangling the contribution of 
reviewer bias from genuine quality differences in low-reputation submissions 
would require controlled study designs or natural experiments exploiting 
reputation threshold effects on Stack Overflow .

%% file: 05-related-work.tex
\section{Related Work}
\label{sec:related-work}

Our work lies at the intersection of collaborative editing analysis, edit rejection modeling, and machine learning–based prioritization under resource constraints. Prior research in this space can be broadly grouped into two categories: (1) Collaborative Editing in Knowledge-Sharing Platforms and (2) Studies of Collaborative Edit Review.

\subsection{Collaborative Editing in Knowledge-Sharing Platforms}

Collaborative editing is an important mechanism for improving the quality of shared knowledge in online communities. Platforms such as Stack Overflow and Wikipedia allow multiple users to modify and refine existing content over time. Prior studies show that collaborative editing can significantly improve the clarity, completeness, and usefulness of shared knowledge artifacts \cite{kittur2008harnessing}. In knowledge-sharing communities, edits often correct mistakes, improve explanations, and update outdated information. For example, Li et al. \cite{isitgood} analyzed five years of Stack Overflow data and found that each substantive edit by another user increased positive votes by 181\% for questions and 119\% for answers. They also found that each edit reduced the later contributions of askers and answerers by no more than 5\%. Jin and Servant \cite{jin2019editsHighlyAnswered} studied edits on highly answered questions. They found that questions that received more answers in a shorter time were often edited by other users, and these edits tended to be small and focused on titles. These results suggest that collaborative editing plays a direct role in improving the perceived quality and usefulness of technical content. 

Stack Overflow and similar systems reward contributors with reputation points and badges, which encourage users to participate in improving existing posts. Empirical studies demonstrate that such incentive mechanisms influence editing behavior. Kittur and Kraut \cite{kittur2008harnessing} found that recognition and coordination mechanisms motivate users to contribute improvements to shared resources, and similar patterns have been observed in Stack Overflow where reputation rewards encourage participation in editing activities. For instance, Munteanu et al.\ \cite{munteanu2008collaborative} studied a webcast platform where users collaboratively corrected transcripts generated by automated speech recognition systems. Their study showed that collaborative corrections significantly reduced transcription errors, demonstrating that community participation can be a cost-effective way to improve system-generated content.

Beyond incentives, several studies have investigated how editing behavior evolves over time. The SOTorrent dataset provides a detailed history of Stack Overflow post versions, which enables researchers to analyze edit timing, edit magnitude, and the co-evolution of code and text segments \cite{baltes2018sotorrent}. Analyses based on this dataset show that many edits occur shortly after a post is created and that edits often modify both code and explanatory text simultaneously. These patterns suggest that edits are commonly used to refine both the technical correctness and the readability of posts. Other studies analyze which parts of posts are edited, such as titles, bodies, or tags, and how editing behavior changes during different stages of a post’s lifecycle \cite{jin2019editsHighlyAnswered, baltes2018sotorrent}. These findings indicate that editing activity is influenced by contextual factors such as question popularity, answer acceptance, and community attention.

Several studies have specifically examined rejected edits and rollback events in Stack Overflow. Wang et al.\ \cite{wang2018reviseAnswers} analyzed rollback revisions and identified common reasons why edits are reverted after being applied. 
Mondal et al. \cite{autopredsaikat1} extended this line of research by developing a taxonomy of rollback rejection reasons and proposing the EditEx tool to assist users in improving suggested edits before submission. Their work also showed that rollback decisions can sometimes be inconsistent \cite{rollbackincnsitency}. Importantly, rollback edits are post-integration reversals applied by high-reputation users, and are fundamentally different from suggested edits, which are rejected before integration by the peer-review queue. Our work is the first to study suggested-edit rejection at scale using the actual SuggestedEdits, ReviewTasks, and ReviewRejectionReasons tables, under the natural class imbalance of the review workflow. These studies highlight the complexity of moderation decisions in collaborative editing environments and emphasize the importance of understanding rejection patterns.

Recent research has also explored the application of machine learning and large language models in Stack Overflow ecosystems \cite{mai2024, soversioning, llmlibrary, bycomforcom, ragreflect, editgenso}. These studies investigate automated edit generation, content transformation, and versioning analysis. Other work has addressed classification and recommendation tasks such as API-related post detection and duplicate question identification \cite{soapi, dupmine}. While these approaches demonstrate the potential of intelligent systems to support community platforms, they primarily focus on content analysis or automated generation rather than understanding the moderation dynamics of suggested edits.

Overall, prior work shows that collaborative editing improves content quality and that incentives such as reputation scores motivate users to participate in editing activities. Existing studies have examined editing behaviors, moderation workflows, and rollback events, and they have proposed tools to assist users in creating better edits. However, there is no prior investigation of suggested-edit decisions before integration. Our work addresses this gap by conducting a large-scale analysis of suggested-edit rejections, developing a refined taxonomy of rejection reasons, and proposing a prioritization framework that predicts rejection probability and supports efficient review decisions within the suggested-edit workflow.

\subsection{Studies of Collaborative Edit Review}

Collaborative editing platforms typically rely on community-based review mechanisms to ensure that edits improve content quality. In systems such as Stack Overflow and Wikipedia, edits submitted by users are often reviewed by other members of the community before being integrated into the content repository. This collaborative review process acts as a quality control layer that prevents harmful or low-quality edits from being accepted. On Stack Overflow, users without full editing privileges must submit suggested edits that are placed in a review queue. Multiple reviewers examine each suggestion and decide whether the edit improves the post or should be rejected \cite{soHelpEditPrivilege, soHelpReviewSuggestedEdits, atwood2011suggestedEdits}. Similar community-driven review mechanisms are used in other collaborative knowledge platforms, where edits must pass through peer review or moderation before they are applied. These mechanisms ensure that collaborative editing remains reliable even when contributions come from a large and diverse group of users.

The review process in collaborative systems is designed to balance openness with quality control. While open participation encourages contributions, review workflows prevent low-quality or harmful changes from degrading the shared knowledge base. However, review tasks can become challenging as platforms grow and the number of submitted edits increases. In Stack Overflow, suggested edits are processed through a shared queue that has a limited capacity. When the queue becomes full, additional edit submissions are temporarily blocked until existing suggestions are reviewed \cite{msoSuggestedQueueFull}. This operational constraint makes efficient review management essential. Reviewers must quickly identify edits that improve content while filtering out edits that provide little value or introduce errors. As a result, understanding how collaborative edit review works and how reviewers prioritize decisions has become an important research topic.

Researchers have explored machine learning approaches to support moderation and review in peer-production platforms. Early work demonstrated that textual signals and readability features can help identify low-quality contributions in Stack Overflow moderation queues \cite{ponzanelli2014lowQualityQueue}. In Wikipedia, automated systems such as ORES provide probabilistic scores that estimate the likelihood that an edit may require review \cite{halfaker2019ores}. These scores help moderators prioritize edits that are more likely to be problematic. Similarly, the STiki system provides vandalism detection by ranking suspicious edits and presenting them to human reviewers in prioritized streams \cite{west2010stiki}. These systems demonstrate that probabilistic models can effectively assist human moderators by focusing attention on higher-risk contributions while still preserving human decision making.

More recent research emphasizes that moderation performance depends not only on prediction accuracy but also on how predictions are integrated into review workflows. Studies such as QUEST model moderation pipelines as queueing systems where routing and prioritization policies influence system efficiency \cite{makhijani2021quest}. In this framework, moderation decisions are affected by factors such as queue size, reviewer availability, and routing strategies. Venire further proposes using probabilistic disagreement signals to determine which items should receive additional review attention \cite{koshy2024venire}. These studies show that probability outputs become most useful when they are integrated into operational policies that guide how items are prioritized within review queues.

Within software engineering processes, similar ideas appear in research on prioritizing code review tasks. Several studies apply learning-to-rank techniques to prioritize pull requests or code changes that require faster decisions \cite{zhao2019ltrPR, vanderveen2015prioritizePR}. Instead of focusing only on classification accuracy, these approaches evaluate ranking performance and early decision effectiveness. The goal is to ensure that limited review resources are directed toward the most impactful or time-sensitive items first. This perspective highlights that ranking and prioritization strategies are critical when human reviewers must process large volumes of submissions.

Overall, prior research demonstrates that collaborative edit review plays a central role in maintaining content quality in peer-production platforms. Existing studies have examined moderation workflows, developed machine learning models for detecting problematic edits. However, there has been limited empirical investigation of suggested-edit rejection behavior and how rejection probabilities can be used to guide prioritization within the Stack Overflow suggested-edit review workflow. Our work addresses this gap by modeling rejection outcomes, developing a refined taxonomy of rejection reasons, and introducing a probability-based prioritization strategy that supports efficient and rejection-aware collaborative edit review.

%% file: 06-threats.tex
\section{Threats to Validity}
\label{sec:threats}
This section discusses threats to the validity of our study following standard guidelines, covering construct, internal, external, and conclusion validity. These threats are important to consider when interpreting our findings and their implications. We also describe steps taken to mitigate these threats.

\subsection{Construct Validity}
Threats to construct validity assesses the extent to which the operational measures accurately capture the intended concepts. Our study assumes that edits accepted by moderators or reviewers are of \textit{higher quality} and rejected edits are of \textit{lower quality}. However, the Stack Overflow review process is known to occasionally reject good edits or accept marginal ones due to reviewer bias, time constraints, or misunderstanding. This introduces noise into our labels, which may affect the validity of our classification models. However, Stack Overflow has strict editing guidelines and each edit in the review queue is evaluated by multiple reviewers, so such cases are expected to be rare relative to the large number of edits analyzed.

To partially mitigate label noise, we exclude suggested-edit conflicts from all analyses. These are not reviewer quality judgments. The conflict label is assigned automatically by the Stack Overflow platform when a concurrent edit is saved to the same post before the suggested edit is processed, making the rejection a system-level race condition rather than a content-based decision~\footnote{https://meta.stackexchange.com/questions/302258}. Regarding user reputation, we reconstructed historical reputation values at edit submission time by aggregating reputation-changing events prior to each edit timestamp. This approximation is consistent with prior work \cite{autopredsaikat1}.

In RQ2, we may not capture every possible nuance of edit rejections. Some categories may overlap (e.g., Minor Format Only Change vs. Undesired Text Modification), or the same edit may arguably fall into multiple categories depending on interpretation. While some edits exhibited more than one plausible rejection reason, each was assigned a single primary label based on the dominant observable change, so borderline cases remain a source of ambiguity. We addressed this through discussion during the annotation process.

\subsection{Internal Validity}
Threats to internal validity refer to bias and errors in conducting our studies. In this paper, we conducted a manual study to understand fine-grained reasons for edit rejections. To avoid bias, each edit was labeled by two of the authors. The level of inter-rater agreement of our qualitative study is high. It is possible that certain rejection categories or user types are underrepresented in our manually validated edits. For example, we may have missed rare or ambiguous rejection scenarios that do not follow the prevalent patterns. To mitigate this, we consider a large number edits for manual study that is statistically representative.

We conduct a quantitative analysis to determine how accepted edits are different from rejected edits. Reviewer decisions may be influenced by factors not fully captured in our study. For example, post topic, post visibility, reviewer workload, time pressure, and familiarity with editing norms may affect acceptance decisions. These latent factors could partially explain the differences observed between accepted and rejected edits. However, we would like to point to the fact the we consider a large number of features that can be captured easily to understand the differences between two groups of edits.

\subsection{External Validity}
Threats to external validity refer to the generalizability of our findings. Our findings are specific to Stack Overflow, a community question-answering website that is specifically designed to discuss programming-related questions. Stack Overflow has a highly structured, reputation-based community with strict editing policies. The results we obtain may not generalize to other collaborative platforms, such as Reddit, Wikipedia, or GitHub where editorial norms differ significantly. However, our study procedures can help to conduct similar studies for other platforms. Future work involving cross-platform validation would further strengthen the robustness and generalizability of the findings. 

Editing norms, reviewer behavior, and platform policies evolve over time. Our dataset reflects a snapshot of suggested edits at a particular time window. Future changes in Stack Overflow’s moderation guidelines, interface, or LLM-assisted editing tools may affect the relevance of our findings. Furthermore, we do not explicitly account for user demographics (e.g., language proficiency, geographic location), which may influence editing behavior and rejection likelihood. This could limit the generalizability of our behavioral insights to different user populations. Future work can focus in this direction. 

It is practically impossible to manually analyze all suggested edits because of their large size and time constraints. Although rare or ambiguous rejection categories may be underrepresented, we mitigate this threat by analyzing a random sample of 690 rejected edits, corresponding to a 99\% confidence level and 5\% margin of error under standard sample-size assumptions.

\subsection{Conclusion Validity}
Conclusion validity refers to the degree to which conclusions are supported by the data. For RQ1, we consider several features across five different dimensions to understand the differences between accepted and rejected edit groups in terms of their mean values. However, this does not necessarily indicate that the underlying populations are different. To mitigate this risk, we evaluate the statistical significance of the observed differences using the Mann–Whitney U test \cite{mann} with Bonferroni correction \cite{abdi2007bonferroni} and Cliff's $\delta$ \cite{cliff}. For RQ3, we compare the performance of different techniques to determine their efficacy in identifying rejected edits. Different datasets or threshold selection can affect the performance of compared techniques. To mitigate this, threshold-based moderation strategies are evaluated under multiple operating points to illustrate trade-offs rather than relying on a single configuration.

%% file: 07-conclusion.tex
\section{Conclusion}
In this study, we conducted a large-scale empirical analysis of 3,888,805  suggested edits submitted to Stack Overflow between 2011 and May 2023. We  investigated how accepted and rejected edits differ, why edits get rejected, and how the review process can be made more efficient. We conduct our study along three research questions (RQs). In RQ1, we compared accepted and rejected edits across 35 features and found statistically significant differences in all 35 features after Bonferroni correction, although most differences had negligible practical effect sizes. The largest differences were observed in user editing history, particularly prior acceptance rate.  To understand why suggested edits get rejected, we manually analyze a statistically representative sample of suggested edits in RQ2. Through manual coding of the sample with high inter-rater agreement ($\kappa = 0.80$), we develop a refined  taxonomy of 17 rejection reasons with concrete examples. Minor formatting changes are the most common rejection reason (35.94\%), while 19.86\% of rejected edits misuse the editing mechanism to add questions, comments, suggestions, or status updates. More critically, 25.51\% introduce incorrect or harmful changes that may undermine post quality or integrity, highlighting the need for careful human review. In RQ3, we evaluated three models for ranking edits by rejection risk so that reviewers can focus on the most problematic submissions. The Feature model provides the strongest trade-off between the proportion of edits sent for human review and the proportion of rejected edits captured and consistently outperformed the embedding model and the EditEx baseline across all decision thresholds. In a complementary zero-shot evaluation of 500 edits, the Feature Model achieved 78.0\% rejected-class recall at $\tau=0.40$, substantially outperforming GPT-4o, Qwen2.5-32B, and Llama-3.3-70B (23.6\%--31.2\%). Our findings from this study can  provide practical guidance for platform designers, reviewers, and  contributors, and encourage future work on fairer and more efficient moderation in collaborative knowledge platforms.

%% file: 09-appendix.tex
\newpage
\section{Detailed Description of Rejection Reasons}
\label{app:rejection_descriptions}
In this section, we provide detailed descriptions and representative examples of the 17 rejection reasons listed in Table~\ref{tab:rejection_taxonomy}. Table~\ref{tab:diff-style-examples} presents an example for each rejection reason. In these examples, added text is shown in \textbf{\emph{bold italics}}, while removed text is shown using \sout{strikeout}.
\paragraph{Minor Format Only Change.}
This reason is used when an edit changes only the appearance of the post. It does not improve the meaning, technical correctness, or usefulness of the content. These edits may change capitalization, spacing, line breaks, or other surface-level formatting. The problem is that the post stays almost the same after the edit. So, the edit does not provide enough value. For example, in PostID 5915892, the edit changes \sout{class} to \emph{Class}. This is only a formatting or capitalization change. It does not improve the question or answer in any meaningful way.

\paragraph{Minor Grammar Only Change.}
This reason is used when an edit fixes small spelling or grammar mistakes, but the improvement is too minor. The edit may be correct, but it still does not make the post much better. SO expects suggested edits to make a clear improvement. A very small grammar fix is often not enough. For example, in PostID 50190912, \sout{is 3 hours} is changed to \emph{are 3 hours}. This correction is grammatically better, but it does not improve the technical value or overall clarity of the post.

\paragraph{Undesired Text Modification.}
This category includes edits that change wording but do not really improve the post. The editor may rewrite words or phrases, but the meaning stays the same. In some cases, the change only makes the text look different, not better. These edits may also change tone or writing style without a strong reason. For example, in PostID 48734717, \sout{Im} is changed to \emph{I'm} several times. The change is correct, but it is still a very small language cleanup. It does not improve the main problem description or solution.

\paragraph{Undesired Code Modification.}
This reason is used when code is changed without a clear benefit. The new code does not improve correctness, readability, or explanation. In some cases, the change may even make the code more confusing. Code edits are more sensitive than text edits because even a small change can affect meaning. For example, in PostID 14124558, part of the code configuration is removed. This removal does not explain anything better and may make the example less complete. So, the edit is rejected.

\paragraph{Spam Link, Keywords.}
This category is used when an edit adds links, keywords, or phrases that are not needed for the post. These additions often look promotional or unrelated. The goal of the edit is not to improve the content, but to insert outside material. This hurts trust and reduces content quality. For example, in PostID 7606910, an external URL is added into the post. The added link is not necessary for improving the explanation. Because of that, the edit is treated as spam-like content.

\paragraph{Meaningless Words.}
This reason is used when an edit adds text that has no clear meaning. The added words may be random characters, nonsense, or irrelevant content. Such edits do not help the reader. Instead, they make the post harder to read and lower its quality. For example, in PostID 18690518, text like \emph{hgfhfhfghfgggrfgdg} is added. This text means nothing. It does not explain the problem or improve the answer. So, it is rejected.

\paragraph{Personal Promotion.}
This category includes edits that promote a product, tool, website, or service. The added material may be loosely related to the topic, but its main purpose is promotion. A suggested edit should improve the post itself, not advertise something. For example, in PostID 810136, external links to a conversion tool and a download page are inserted. These links are promotional. They do not improve the original technical explanation. Therefore, the edit is rejected as personal promotion.
\paragraph{Post Defacement.}
This reason is used when an edit damages the post. The edit may remove useful content, replace it with empty text, or make the post much less useful. This is more serious than a low-value edit because it directly harms the post. For example, in PostID 22939493, useful text and code are removed, and the post is replaced with \emph{Edit: .............}. This change destroys the value of the answer. So, it is considered post defacement.
\paragraph{Rude or Abusive Words.}
This category is used when an edit adds insulting, offensive, or abusive language. These edits clearly violate community rules. They do not improve the post and may create a hostile environment. For example, in PostID 3269169, a polite sentence is changed into \emph{You do not need bravery. Stop being a brainwashed droid.} This added text is rude and insulting. It is rejected because it harms both the post and the community tone.

\paragraph{Comment on Post.}
This category is used when an edit adds discussion-style text. The added content may be a note, message, announcement, or explanation directed at readers or future contributors. That kind of text belongs in comments, not in the post body. For example, in PostID 15488099, a bounty-related announcement is added for future answers. This is discussion content. It does not improve the original question or answer. So, it is classified as a comment on the post.

\paragraph{Status Update.}
This reason is used when an edit adds temporary or meta-level information. The added text may explain the edit itself, mention version status, or include a note that is not part of the real content. These edits do not improve the long-term value of the post. For example, in PostID 3865390, the edit adds \emph{WARNING. This long line is only here to satisfy minimum edit 6chars requirement. Please remove it.} This text is only about the edit process. It does not improve the technical content, so it is rejected.

\paragraph{Invalid Correction.}
This category is used when the editor tries to fix the post, but the result is still poor or partly wrong. The intent may be good, but the correction is weak, incomplete, or inconsistent. The new version may still have grammar problems, awkward wording, or broken sentence flow. For example, in PostID 14180518, parts of the text are rewritten, such as changing \sout{Gridview then} to \emph{gridview}, while other parts are only partly corrected. The result is not clear and does not fully solve the language problem. So, the edit is rejected as an invalid correction.

\paragraph{Incorrect Syntax.}
This reason is used when an edit adds broken code or invalid syntax. This is a serious problem because readers may trust and copy the code from the post. If the added code is syntactically wrong, the edit makes the post less reliable. For example, in PostID 228973, code like \emph{fclose(datei) strtok('hej',string)} is inserted. This code is malformed and does not fit correctly into the example. So, the edit is rejected under incorrect syntax.

\paragraph{Alternate Suggestion.}
This category includes edits that add a different solution instead of improving the existing one. The added solution may even be technically valid, but it still does not belong in an edit. Suggested edits should refine the current answer, not attach a new answer to it. For example, in PostID 3172358, a different implementation using extra conditional logic is added. This is a new suggestion, not an improvement of the existing solution. Therefore, it is rejected.

\paragraph{Removal of Essential Context.}
This reason is used when an edit removes important content and weakens the purpose of the post. The removed part may be code, explanation, context, or an example that is necessary for understanding the post. After the edit, the post may become incomplete or much less useful. For example, in PostID 11684579, useful code and explanation are removed. After that, the post no longer serves its original purpose well. So, the edit is rejected.

\begin{table}[H]
\centering
\caption{Examples for newly defined rejection reasons (added text presented in \textbf{bold \emph{italics}}, removed text in \sout{strikeout}).}
\label{tab:diff-style-examples}
\footnotesize
\setlength{\tabcolsep}{6pt}
\renewcommand{\arraystretch}{1.2}
\begin{tabular}{|p{2.5cm}|p{9.6cm}|}
\hline
\textbf{Our Defined Reasons} & \textbf{Edit Submitted (PostID)} \\
\hline

Minor Format Only Change
&
Assuming that I have a class named \sout{class} \textbf{\emph{Class}}, 
And I would like to make a new ArrayList \emph{so} that it's values will be of type \sout{class} \textbf{\emph{Class}}. \newline 
(PostID: \href{https://stackoverflow.com/q/5915892}{5915892})
\\
\hline

Minor Grammar Only Change
&
I want to reach screen time of every application that's working on android system. 
For example, there \sout{is 3 hour} \textbf{\emph{are 3 hours}} of facebook working time but \sout{1.2 hour} \textbf{\emph{1.2 hours}} of screen time. \newline
(PostID: \href{https://stackoverflow.com/q/50190912}{50190912})
\\
\hline
Undesired Text Modification
&
\sout{Im} \textbf{\emph{I'm}} pulling my hair out trying to center this page \ldots but I cant get the page to center and \sout{im} \textbf{\emph{I'm}} not sure why it will not center \ldots unless \sout{im} \textbf{\emph{I'm}} using the wrong element, then \sout{im} \textbf{\emph{I'm}} not sure what \sout{im} \textbf{\emph{I'm}} doing wrong. 

(PostID: \href{https://stackoverflow.com/q/48734717}{48734717})
\\
\hline

Undesired Code Modification
&
\texttt{\$this->addElement(} 
\texttt{"select","selectName",} \newline
\sout{\texttt{array("multiOptions"=>array("one","two","three"),}} \newline
\texttt{"disable"=>array(0,1))} \newline
(PostID: \href{https://stackoverflow.com/q/14124558}{14124558})
\\
\hline

Spam link, keywords
&

you need to specify FROM ex:

\sout{SELECT f1id, owner, name FROM friendlist WHERE owner=me()}

\textbf{\emph{String fqlurl="https://api.facebook.com/method/fql.query?}}
\newline
\textbf{\emph{SELECT f1id, owner, name FROM friendlist WHERE owner=me()";}}

(PostID: \href{https://stackoverflow.com/q/7606910}{7606910})
\\
\hline
Meaningless words 

&
Parse, UITableView and reload data 
\textbf{\emph{hgfhfhfghfgggrfgdg gdfgdfg fgdgfdgdfg gdgfsg}}

(PostID: \href{https://stackoverflow.com/q/18690518}{18690518})
\\
\hline

Personal promotion 

&
(Explains OST to PST conversion steps.) 
\textbf{\emph{http://www.osttopstconversion.us/ -- Please visit us}}
\textbf{\emph{http://www.systoolsdl.com/get.php?id=50 -- }}\newline 
\textbf{\emph{That is download link}} 

(PostID: \href{https://stackoverflow.com/q/810136}{810136})
\\
\hline




Post Defacement
&
\sout{Please try this code. No need for the transformArray or Collection outlets or the transform variables that you created. Please check it gives you the desired result.} \newline
\sout{\texttt{(IBAction) buttonsHighlighted:(id)sender \{}}\newline
... \newline
\sout{\texttt{    if(btn.isHighlighted) \{ scale = 1.05; \}}} \newline
\textbf {\emph{Edit: .............}}\newline
(PostID: \href{https://stackoverflow.com/q/22939493}{22939493})
\\
\hline

Rude or abusive words 

&
I actually haven't been brave enough to write my pics in assembly. \sout{Good luck!!} 
\textbf{\emph{You do not need bravery. Stop being a brainwashed droid.}}
(PostID: \href{https://stackoverflow.com/q/3269169}{3269169})
\\
\hline


Comment on post 

&
(Question text / numbered list continues.) 
\sout{3. Does it have deeper consequences?}
\textbf{\emph{Announcement by bounty starter:}} 
\textbf{\emph{For new coming answers, please consider explain... \ldots}} 
\textbf{\emph{(this message will be removed after the bounty is awarded)}}

(PostID: \href{https://stackoverflow.com/q/15488099}{15488099})
\\
\hline

Status Update
&
Is there a command in \sout{WFF} \textbf{\emph{WPF}} to close the application from context menu? \newline
\textbf{\emph{WARNING. This long line is only here to satisfy minimum edit 6chars requirement.}} \newline
\textbf{\emph{ Please remove it.} }\newline
(PostID: \href{https://stackoverflow.com/q/3865390}{3865390})
\\
\hline

\end{tabular}

\end{table}

\begin{table}[H]
\centering
\footnotesize
\setlength{\tabcolsep}{6pt}
\renewcommand{\arraystretch}{1.25}
\begin{tabular}{|p{2.5cm}|p{9.6cm}|}
\hline
\textbf{Our Defined Reasons} &  \textbf{Edit Submitted (PostID)} \\
\hline
% ...


Invalid correction

&
Click "Accept" button in \sout{Gridview then} \textbf{\emph{gridview}} and move \sout{the record in data from one gridview to another Gridview in different page} \textbf{\emph{data from one gridview to another gridview in a different page}}. \sout{I} \textbf{\emph{i}} want a gridview that having "ACCEPT" button then after clicking it. \sout{I} \textbf{\emph{i}} want the accepted record to be moved or transferred in another gridview which is located in another page \sout{?}. \sout{I} \textbf{\emph{i}} already have two gridviews: \sout{Gridview1 named PendingRecordsGridview. Gridview2 named AcceptedRecordsGridview} \textbf{\emph{Gridview1 named PendingRecordsGridview and Gridview2 named AcceptedRecordsGridview}}.
(PostID: \href{https://stackoverflow.com/q/14180518}{14180518})
\\
\hline
Incorrect syntax 

&
Open "myfile.csv" For Input As 1 \newline
...
V = Split(Txt, ",") \newline
\textbf{{\emph{fclose(datei) strtok('hej',string)}}}

(PostID: \href{https://stackoverflow.com/q/228973}{228973})
\\
\hline
Alternate suggestion 

&
You can use the \texttt{gmdate()} function: \newline
   \texttt{echo gmdate("H:i:s", 685);}
\textbf{\emph{if you need more control:} }
\textbf{\emph{\$time = 3800;} }
\textbf{\emph{\$time = (\$time < 3600) ? gmdate("i:s", \$time) : gmdate("H:i:s", \$time);}}

(PostID: \href{https://stackoverflow.com/q/3172358}{3172358})
\\
\hline
Removal of essential context 

&
Unexpected call to method or property access on IE8 in jQuery 
\emph{Error occuring while opening http://garmonia-znakomstva.ru/} 
\sout{(full working code example and explanation remain here)} 
\textbf{\emph{(code/content removed such that the original example’s purpose is lost)}}

(PostID: \href{https://stackoverflow.com/q/11684579}{11684579})
\\
\hline
Question on post

&
Any help in getting this sorted out would be much appreciated \ldots 
\textbf{\emph{What resources are you using? }}

\textbf{\emph{Can you provide more info on the specific app, services, 
}} \newline
\textbf{\emph{ 
and resources you are using and what is not appearing in the new portal?}}
\textbf{\emph{... Is this a web app using cloud services?}}

(PostID: \href{https://stackoverflow.com/q/38507190}{38507190})
\\
\hline

\end{tabular}
\end{table}

\paragraph{Question on Post.}
This category is used when an edit adds a new question inside the post. Instead of improving the existing content, the editor starts asking for more information. This is not the role of a suggested edit. These kinds of additions belong in comments, not in the post text. For example, in PostID 38507190, the editor adds questions such as \emph{What resources are you using?} and asks for more details. These are follow-up questions, not improvements. So, the edit is rejected.

\paragraph{Other.}
This is a fallback category. It is used when a rejected edit does not clearly match any of the refined reasons. Sometimes the edit is too unclear, too mixed, or missing enough context to classify it with confidence. Instead of forcing it into the wrong category, it is placed in Other. This helps keep the taxonomy more accurate and reliable.

\paragraph{Overall Interpretation.}
These rejection reasons show that rejected edits are not all the same. Some are low-value edits, such as formatting-only changes or very small grammar fixes. Others are clearly harmful, such as spam, abusive text, broken code, or content removal. This difference is important. It shows that many rejected edits are not malicious, but they still fail because they do not improve the post enough. The taxonomy helps us better understand editor behavior and gives a clearer basis for moderation analysis and automated rejection prediction.

%% file: references.bib
@article{cliff,
  author  = {Cliff, Norman},
  title   = {Dominance Statistics: Ordinal Analyses to Answer Ordinal Questions},
  journal = {Psychological Bulletin},
  year    = {1993},
  volume  = {114},
  number  = {3},
  pages   = {494--509},
  doi     = {10.1037/0033-2909.114.3.494}
}

@article{romano,
  title={Using Cliff’s delta as a non-parametric effect size measure: an accessible web app and R tutorial},
  author={Meissel, Kane and Yao, Esther S},
  journal={Practical Assessment, Research, and Evaluation},
  volume={29},
  number={1},
  year={2024},
  publisher={University of Massachusetts Amherst Libraries}
}

@inproceedings{munteanu2008collaborative,
  title={Collaborative editing for improved usefulness and usability of transcript-enhanced webcasts},
  author={Munteanu, Cosmin and Baecker, Ron and Penn, Gerald},
  booktitle={Proceedings of the SIGCHI conference on human factors in computing systems},
  pages={373--382},
  year={2008}
}

@misc{openai2024gpt4,
      title={GPT-4o System Card}, 
      author={OpenAI Team},
      year={2024},
      eprint={2410.21276},
      archivePrefix={arXiv},
      primaryClass={cs.CL},
      url={https://arxiv.org/abs/2410.21276}, 
}

@inproceedings{yang2024large,
author = {Yang, Aidan Z. H. and Le Goues, Claire and Martins, Ruben and Hellendoorn, Vincent},
title = {Large Language Models for Test-Free Fault Localization},
year = {2024},
isbn = {9798400702174},
publisher = {Association for Computing Machinery},
address = {New York, NY, USA},
doi = {10.1145/3597503.3623342},
articleno = {17},
numpages = {12},
location = {Lisbon, Portugal},
series = {ICSE '24}
}

@misc{sun2024llm4vuln,
      title={LLM4Vuln: A Unified Evaluation Framework for Decoupling and Enhancing LLMs' Vulnerability Reasoning}, 
      author={Yuqiang Sun and Daoyuan Wu and Yue Xue and Han Liu and Wei Ma and Lyuye Zhang and Yang Liu and Yingjiu Li},
      year={2025},
      eprint={2401.16185},
      archivePrefix={arXiv},
      primaryClass={cs.CR},
      url={https://arxiv.org/abs/2401.16185}, 
}

@inproceedings{niu2024empirical,
author = {Rasnayaka, Sanka and Wang, Guanlin and Shariffdeen, Ridwan and Iyer, Ganesh Neelakanta},
title = {An Empirical Study on Usage and Perceptions of LLMs in a Software Engineering Project},
year = {2024},
isbn = {9798400705793},
publisher = {Association for Computing Machinery},
address = {New York, NY, USA},
doi = {10.1145/3643795.3648379},
booktitle = {Proceedings of the 1st International Workshop on Large Language Models for Code},
pages = {111–118},
numpages = {8},
location = {Lisbon, Portugal},
series = {LLM4Code '24}
}

@INPROCEEDINGS{li2024codereviewer,
  author={Cihan, Umut and Haratian, Vahid and {\.I}{\c{c}}{\"o}z, Arda and Gül, Mert Kaan and Devran, {\"O}mercan and Bayendur, Emircan Furkan and Uçar, Baykal Mehmet and Tüzün, Eray},
  booktitle={2025 IEEE/ACM 47th International Conference on Software Engineering: Software Engineering in Practice (ICSE-SEIP)}, 
  title={Automated Code Review in Practice}, 
  year={2025},
  volume={},
  number={},
  pages={425-436},
  doi={10.1109/ICSE-SEIP66354.2025.00043}}

@misc{qwen2024,
      title={Qwen2.5 Technical Report}, 
      author={Qwen Team},
      year={2025},
      eprint={2412.15115},
      archivePrefix={arXiv},
      primaryClass={cs.CL},
      url={https://arxiv.org/abs/2412.15115}, 
}

@article{dubey2024llama,
  title={The Llama 3 Herd of Models},
  author={Grattafiori, Aaron and Dubey, Abhimanyu and Jauhri, Abhinav and Pandey, Abhinav and Kadian, Abhishek and Al-Dahle, Ahmad and Letman, Aiesha and Mathur, Akhil and Schelten, Alan and Yang, Amy and Fan, Angela and others},
  journal={arXiv preprint arXiv:2407.21783},
  year={2024},
  url={https://arxiv.org/abs/2407.21783}
}

@inproceedings{white2023chatgpt,
author = {White, Jules and Fu, Quchen and Hays, Sam and Sandborn, Michael and Olea, Carlos and Gilbert, Henry and Elnashar, Ashraf and Spencer-Smith, Jesse and Schmidt, Douglas C.},
title = {A Prompt Pattern Catalog to Enhance Prompt Engineering with ChatGPT},
year = {2023},
isbn = {9781941652190},
publisher = {The Hillside Group},
address = {USA},
booktitle = {Proceedings of the 30th Conference on Pattern Languages of Programs},
articleno = {5},
numpages = {31},
location = {Monticello, IL, USA},
series = {PLoP '23}
}

@software{replication,
  author = {Mehedi Hasan Shanto},
  title = {SO-Suggested Edits},
  url = {https://github.com/mehedisresearch/SO-Manual-Study},
  version = {1.0.0},
  date = {2026-04-30}
}

@article{bycomforcom,
author = {Chen, Chunyang and Xing, Zhenchang and Liu, Yang},
title = {By the Community \& For the Community: A Deep Learning Approach to Assist Collaborative Editing in Q\&A Sites},
year = {2017},
issue_date = {November 2017},
publisher = {Association for Computing Machinery},
address = {New York, NY, USA},
volume = {1},
number = {CSCW},
url = {https://doi.org/10.1145/3134667},
doi = {10.1145/3134667},
journal = {Proc. ACM Hum.-Comput. Interact.},
month = dec,
articleno = {32},
numpages = {21}
}

@inproceedings{dupmine,
author = {Ahasanuzzaman, Muhammad and Asaduzzaman, Muhammad and Roy, Chanchal K. and Schneider, Kevin A.},
title = {Mining duplicate questions in stack overflow},
year = {2016},
isbn = {9781450341868},
publisher = {Association for Computing Machinery},
address = {New York, NY, USA},
doi = {10.1145/2901739.2901770},
booktitle = {Proceedings of the 13th International Conference on Mining Software Repositories},
pages = {402–412},
numpages = {11},
location = {Austin, Texas},
series = {MSR '16}
}

@INPROCEEDINGS{soapi,
  author={Ahasanuzzaman, Md and Asaduzzaman, Muhammad and Roy, Chanchal K. and Schneider, Kevin A.},
  booktitle={2018 IEEE 25th International Conference on Software Analysis, Evolution and Reengineering (SANER)}, 
  title={Classifying stack overflow posts on API issues}, 
  year={2018},
  volume={},
  number={},
  pages={244-254},
  doi={10.1109/SANER.2018.8330213}}

@misc{llmlibrary,
      title={How Robust are LLM-Generated Library Imports? An Empirical Study using Stack Overflow}, 
      author={Jasmine Latendresse and SayedHassan Khatoonabadi and Emad Shihab},
      year={2025},
      eprint={2507.10818},
      archivePrefix={arXiv},
      primaryClass={cs.SE},
      url={https://arxiv.org/abs/2507.10818}, 
}

@inproceedings{soversioning,
author = {Chen, Yiyun and Chen, Fuxiang and Kim, Mijung and Fard, Fatemeh},
title = {Studying Versioning in Stack Overflow},
year = {2024},
isbn = {9798400712487},
publisher = {Association for Computing Machinery},
address = {New York, NY, USA},
doi = {10.1145/3691620.3695326},
booktitle = {Proceedings of the 39th IEEE/ACM International Conference on Automated Software Engineering},
pages = {2456–2457},
numpages = {2},
location = {Sacramento, CA, USA},
series = {ASE '24}
}

@article{ragreflect,
  title={RAG-Reflect: Agentic Retrieval-Augmented Generation with Reflections for Comment-Driven Code Maintenance on Stack Overflow},
  author={Shanto, Mehedi Hasan and Asaduzzaman, Muhammad and Ngom, Alioune},
  journal={arXiv preprint arXiv:2604.22217},
  year={2026}
}

@INPROCEEDINGS{mai2024,
  author={Mai, Yubo and Gao, Zhipeng and Wang, Haoye and Bi, Tingting and Hu, Xing and Xia, Xin and Sun, Jianling},
  booktitle={2025 IEEE/ACM 47th International Conference on Software Engineering (ICSE)}, 
  title={Towards Better Answers: Automated Stack Overflow Post Updating}, 
  year={2025},
  volume={},
  number={},
  pages={591-603},
  doi={10.1109/ICSE55347.2025.00024}}

@misc{qwen2,
      title={Qwen2 Technical Report}, 
      author={Qwen Team},
      year={2024},
      eprint={2407.10671},
      archivePrefix={arXiv},
      primaryClass={cs.CL},
      url={https://arxiv.org/abs/2407.10671}, 
}

@inproceedings{Kabir_2024,
author = {Kabir, Samia and Udo-Imeh, David N. and Kou, Bonan and Zhang, Tianyi},
title = {Is Stack Overflow Obsolete? An Empirical Study of the Characteristics of ChatGPT Answers to Stack Overflow Questions},
year = {2024},
isbn = {9798400703300},
publisher = {Association for Computing Machinery},
address = {New York, NY, USA},
doi = {10.1145/3613904.3642596},
booktitle = {Proceedings of the 2024 CHI Conference on Human Factors in Computing Systems},
articleno = {935},
numpages = {17},
location = {Honolulu, HI, USA},
series = {CHI '24}
}

@book{svm,
  title={Support vector machines},
  author={Steinwart, Ingo and Christmann, Andreas},
  year={2008},
  publisher={Springer Science \& Business Media}
}

@inproceedings{knn,
  title={KNN model-based approach in classification},
  author={Guo, Gongde and Wang, Hui and Bell, David and Bi, Yaxin and Greer, Kieran},
  booktitle={OTM Confederated International Conferences" On the Move to Meaningful Internet Systems"},
  pages={986--996},
  year={2003},
  organization={Springer}
}

@article{rf,
  title={Random forests},
  author={Breiman, Leo},
  journal={Machine learning},
  volume={45},
  number={1},
  pages={5--32},
  year={2001},
  publisher={Springer}
}

@article{scikit-learn,
author = {Pedregosa, Fabian and Varoquaux, Ga{\"e}l and Gramfort, Alexandre and Michel, Vincent and Thirion, Bertrand and Grisel, Olivier and Blondel, Mathieu and Prettenhofer, Peter and Weiss, Ron and Dubourg, Vincent and Vanderplas, Jake and Passos, Alexandre and Cournapeau, David and Brucher, Matthieu and Perrot, Matthieu and Duchesnay, {\'E}douard},
title = {Scikit-learn: Machine Learning in Python},
year = {2011},
issue_date = {2/1/2011},
publisher = {JMLR.org},
volume = {12},
number = {85},
issn = {1532-4435},
journal = {J. Mach. Learn. Res.},
month = nov,
pages = {2825–2830},
numpages = {6}
}

@inproceedings{xgboost,
  author    = {Chen, Tianqi and Guestrin, Carlos},
  title     = {XGBoost: A Scalable Tree Boosting System},
  booktitle = {Proceedings of the 22nd ACM SIGKDD International Conference on Knowledge Discovery and Data Mining (KDD)},
  year      = {2016},
  pages     = {785--794}
}

@article{mann,
  author    = {Mann, Henry B. and Whitney, Donald R.},
  title     = {On a test of whether one of two random variables is stochastically larger than the other},
  journal   = {The Annals of Mathematical Statistics},
  volume    = {18},
  number    = {1},
  pages     = {50--60},
  year      = {1947},
  publisher = {Institute of Mathematical Statistics},
  doi       = {10.1214/aoms/1177730491},
  url       = {https://doi.org}
}

@inproceedings{baltes2018sotorrent,
  author    = {Sebastian Baltes and Lorik Dumani and Christoph Treude and Stephan Diehl},
  title     = {{SOTorrent: Reconstructing and Analyzing the Evolution of Stack Overflow Posts}},
  booktitle = {Proceedings of the 15th International Conference on Mining Software Repositories (MSR)},
  year      = {2018},
  doi       = {10.1145/3196398.3196430}
}

@inproceedings{jin2019editsHighlyAnswered,
  author    = {Xianhao Jin and Francisco Servant},
  title     = {{What Edits Are Done on The Highly Answered Questions in Stack Overflow? An Empirical Study}},
  booktitle = {MSR 2019 Mining Challenge Track},
  year      = {2019}
}

@ARTICLE{wang2018reviseAnswers,
author={Wang, Shaowei and Chen, Tse-Hsun and Hassan, Ahmed E.},
journal={ IEEE Transactions on Software Engineering },
title={{ How Do Users Revise Answers on Technical Q\&A Websites? A Case Study on Stack Overflow }},
year={2020},
volume={46},
number={09},
ISSN={1939-3520},
pages={1024-1038},
doi={10.1109/TSE.2018.2874470},
url = {https://doi.ieeecomputersociety.org/10.1109/TSE.2018.2874470},
publisher={IEEE Computer Society},
address={Los Alamitos, CA, USA},
month=sep}

@inproceedings{ponzanelli2014lowQualityQueue,
author = {Ponzanelli, Luca and Mocci, Andrea and Bacchelli, Alberto and Lanza, Michele and Fullerton, David},
title = {Improving Low Quality Stack Overflow Post Detection},
year = {2014},
isbn = {9781479961467},
publisher = {IEEE Computer Society},
address = {USA},
doi = {10.1109/ICSME.2014.90},
booktitle = {Proceedings of the 2014 IEEE International Conference on Software Maintenance and Evolution},
pages = {541–544},
numpages = {4},
series = {ICSME '14}
}

@article{makhijani2021quest,
  author  = {Rahul Makhijani and Parikshit Shah and Vashist Avadhanula and Caner Gocmen and Nicol{\'a}s E. Stier{-}Moses and Juli{\'a}n Mestre},
  title   = {{QUEST: Queue Simulation for Content Moderation at Scale}},
  journal = {arXiv preprint arXiv:2103.16816},
  year    = {2021}
}

@article{koshy2024venire,
author = {Koshy, Vinay and Choi, Frederick and Chiang, Yi-Shyuan and Sundaram, Hari and Chandrasekharan, Eshwar and Karahalios, Karrie},
title = {Venire: A Machine Learning-Guided Panel Review System for Community Content Moderation},
year = {2025},
issue_date = {November 2025},
publisher = {Association for Computing Machinery},
address = {New York, NY, USA},
volume = {9},
number = {7},
url = {https://doi.org/10.1145/3757699},
doi = {10.1145/3757699},
journal = {Proc. ACM Hum.-Comput. Interact.},
month = oct,
articleno = {CSCW518},
numpages = {35}
}

@article{halfaker2019ores,
author = {Halfaker, Aaron and Geiger, R. Stuart},
title = {ORES: Lowering Barriers with Participatory Machine Learning in Wikipedia},
year = {2020},
issue_date = {October 2020},
publisher = {Association for Computing Machinery},
address = {New York, NY, USA},
volume = {4},
number = {CSCW2},
url = {https://doi.org/10.1145/3415219},
doi = {10.1145/3415219},
journal = {Proc. ACM Hum.-Comput. Interact.},
month = oct,
articleno = {148},
numpages = {37}
}

@article{zhao2019ltrPR,
author = {Zhao, Guoliang and Costa, Daniel Alencar and Zou, Ying},
title = {Improving the pull requests review process using learning-to-rank algorithms},
year = {2019},
issue_date = {August    2019},
publisher = {Kluwer Academic Publishers},
address = {USA},
volume = {24},
number = {4},
issn = {1382-3256},
url = {https://doi.org/10.1007/s10664-019-09696-8},
doi = {10.1007/s10664-019-09696-8},
journal = {Empirical Softw. Engg.},
month = aug,
pages = {2140–2170},
numpages = {31}
}

@INPROCEEDINGS{vanderveen2015prioritizePR,
  author={van der Veen, Erik and Gousios, Georgios and Zaidman, Andy},
  booktitle={2015 IEEE/ACM 12th Working Conference on Mining Software Repositories}, 
  title={Automatically Prioritizing Pull Requests}, 
  year={2015},
  volume={},
  number={},
  pages={357-361},
  doi={10.1109/MSR.2015.40}}

@inproceedings{west2010stiki,
author = {West, Andrew G. and Kannan, Sampath and Lee, Insup},
title = {STiki: an anti-vandalism tool for Wikipedia using spatio-temporal analysis of revision metadata},
year = {2010},
isbn = {9781450300568},
publisher = {ACM},
address = {New York, NY, USA},
doi = {10.1145/1832772.1832814},
booktitle = {Proceedings of the 6th International Symposium on Wikis and Open Collaboration},
articleno = {32},
numpages = {2},
location = {Gdansk, Poland},
series = {WikiSym '10}
}

@misc{soHelpEditPrivilege,
  title        = {Stack Overflow Help Center: Privileges -- Edit posts},
  howpublished = {\url{https://stackoverflow.com/help/privileges/edit}},
note = {Accessed: 2026-05-07}

}

@misc{soHelpReviewSuggestedEdits,
  title        = {Stack Overflow Help Center: How do I use the Suggested edits queue?},
  howpublished = {\url{https://stackoverflow.com/help/review-suggested-edits}},
    note = {Accessed: 2025-05-07}

}

@inproceedings{isitgood,
author = {Li, Guo and Zhu, Haiyi and Lu, Tun and Ding, Xianghua and Gu, Ning},
title = {Is It Good to Be Like Wikipedia? Exploring the Trade-offs of Introducing Collaborative Editing Model to Q\&A Sites},
year = {2015},
isbn = {9781450329224},
publisher = {Association for Computing Machinery},
address = {New York, NY, USA},
doi = {10.1145/2675133.2675155},
pages = {1080–1091},
numpages = {12},
location = {Vancouver, BC, Canada},
series = {CSCW '15}
}

@misc{atwood2011suggestedEdits,
  title        = {Stack Overflow Blog: Suggested Edits and Edit Review},
  author       = {Jeff Atwood},
  year         = {2011},
  howpublished = {\url{https://stackoverflow.blog/2011/02/05/suggested-edits-and-edit-review/}}
}

@misc{msoSuggestedQueueFull,
  title        = {Meta Stack Overflow: Suggested edit queue is full},
  howpublished = {\url{https://meta.stackoverflow.com/questions/345125/suggested-edit-queue-is-full}},
  note = {Accessed: 2026-05-07}

}

@inproceedings{mamykina2011design,
author = {Mamykina, Lena and Manoim, Bella and Mittal, Manas and Hripcsak, George and Hartmann, Bj{\"o}rn},
title = {Design lessons from the fastest \&a site in the west},
year = {2011},
isbn = {9781450302289},
publisher = {Association for Computing Machinery},
address = {New York, NY, USA},
doi = {10.1145/1978942.1979366},
booktitle = {Proceedings of the SIGCHI Conference on Human Factors in Computing Systems},
pages = {2857–2866},
numpages = {10},
location = {Vancouver, BC, Canada},
series = {CHI '11}
}

@inproceedings{anderson2012discovering,
author = {Anderson, Ashton and Huttenlocher, Daniel and Kleinberg, Jon and Leskovec, Jure},
title = {Discovering value from community activity on focused question answering sites: a case study of stack overflow},
year = {2012},
isbn = {9781450314626},
publisher = {Association for Computing Machinery},
address = {New York, NY, USA},
doi = {10.1145/2339530.2339665},
booktitle = {Proceedings of the 18th ACM SIGKDD International Conference on Knowledge Discovery and Data Mining},
pages = {850–858},
numpages = {9},
location = {Beijing, China},
series = {KDD '12}
}

@inproceedings{arpon,
author = {Islam, Md Anaytul and Asaduzzman, Muhammad and Wang, Shaowei},
title = {On the Executability of R Markdown Files},
year = {2024},
isbn = {9798400705878},
publisher = {Association for Computing Machinery},
address = {New York, NY, USA},
doi = {10.1145/3643991.3644931},
booktitle = {Proceedings of the 21st International Conference on Mining Software Repositories},
pages = {254–264},
numpages = {11},
location = {Lisbon, Portugal},
series = {MSR '24}
}

@ARTICLE{obsoluteanswer,
author={Zhang, Haoxiang and Wang, Shaowei and Chen, Tse-Hsun and Zou, Ying and Hassan, Ahmed E.},
journal={ IEEE Transactions on Software Engineering },
title={{ An Empirical Study of Obsolete Answers on Stack Overflow }},
year={2021},
volume={47},
number={4},
ISSN={1939-3520},
pages={850-862},
doi={10.1109/TSE.2019.2906315},
publisher={IEEE Computer Society},
address={Los Alamitos, CA, USA},
month=apr}

@inbook{codechange,
author = {Opu, Md Nahidul Islam and Wang, Shaowei and Chowdhury, Shaiful},
title = {LLM-Based Detection of Tangled Code Changes for Higher-Quality Method-Level Bug Datasets},
year = {2026},
isbn = {9798400724749},
publisher = {Association for Computing Machinery},
address = {New York, NY, USA},
doi={10.1145/3793302.3793351},
booktitle = {Proceedings of the 23rd International Conference on Mining Software Repositories},
pages = {199–211},
numpages = {13}
}

@inproceedings{llm4code,
author = {Nam, Daye and Macvean, Andrew and Hellendoorn, Vincent and Vasilescu, Bogdan and Myers, Brad},
title = {Using an LLM to Help With Code Understanding},
year = {2024},
isbn = {9798400702174},
publisher = {Association for Computing Machinery},
address = {New York, NY, USA},

doi = {10.1145/3597503.3639187},
booktitle = {Proceedings of the IEEE/ACM 46th International Conference on Software Engineering},
articleno = {97},
numpages = {13},
location = {Lisbon, Portugal},
series = {ICSE '24}
}

@misc{llm2vec,
      title={LLM2Vec: Large Language Models Are Secretly Powerful Text Encoders}, 
      author={Parishad BehnamGhader and Vaibhav Adlakha and Marius Mosbach and Dzmitry Bahdanau and Nicolas Chapados and Siva Reddy},
      year={2024},
      eprint={2404.05961},
      archivePrefix={arXiv},
      primaryClass={cs.CL},
      url={https://arxiv.org/abs/2404.05961}, 
}

@INPROCEEDINGS{rollbackincnsitency,
  author={Mondal, Saikat and Uddin, Gias and Roy, Chanchal K.},
  booktitle={2021 IEEE/ACM 18th International Conference on Mining Software Repositories (MSR)}, 
  title={Rollback Edit Inconsistencies in Developer Forum}, 
  year={2021},
  volume={},
  number={},
  pages={380-391},
  doi={10.1109/MSR52588.2021.00050}}

@article{autopredsaikat1,
author = {Mondal, Saikat and Uddin, Gias and Roy, Chanchal},
title = {Automatic prediction of rejected edits in Stack Overflow},
year = {2023},
issue_date = {Jan 2023},
publisher = {Kluwer Academic Publishers},
address = {USA},
volume = {28},
number = {1},
issn = {1382-3256},
url = {https://doi.org/10.1007/s10664-022-10242-2},
doi = {10.1007/s10664-022-10242-2},
journal = {Empirical Softw. Engg.},
month = jan,
articleno = {9},
numpages = {43},}

@misc{so-edit-policy,
  title = {How do suggested edits work?},
  howpublished = {\url{https://meta.stackexchange.com/questions/76251/how-do-suggested-edits-work}},
  note = {Accessed: 2025-08-07}
}

@misc{so_edit_reward,
  title = {Editing is essential: new badges and review enhancements},
  howpublished = {\url{https://stackoverflow.blog/2014/10/07/new-editing-badges-and-enhancements-to-suggested-edits/}},
  note = {Accessed: 2025-08-07}
}

@inproceedings{kittur2008harnessing,
author = {Kittur, Aniket and Kraut, Robert E.},
title = {Harnessing the wisdom of crowds in wikipedia: quality through coordination},
year = {2008},
isbn = {9781605580074},
publisher = {Association for Computing Machinery},
address = {New York, NY, USA},
doi = {10.1145/1460563.1460572},
pages = {37–46},
numpages = {10},
location = {San Diego, CA, USA},
series = {CSCW '08}
}

@article{abdi2007bonferroni,
  title={Bonferroni and {\v{S}}id{\'a}k corrections for multiple comparisons},
  author={Abdi, Herv{\'e}},
  journal={Encyclopedia of measurement and statistics},
  volume={3},
  number={01},
  pages={2007},
  year={2007},
  publisher={Sage, Thousand Oaks, CA}
}

@misc{alibabanlp2024gteqwen2,
author = {{Alibaba-NLP}},
title = {{gte-Qwen2-7B-instruct} Model Card},
year = {2024},
howpublished = {Hugging Face},
url = {https://huggingface.co/Alibaba-NLP/gte-Qwen2-7B-instruct},
note = {Accessed 2026-11-04}
}

@misc{meta2024llama33,
author = {{Meta}},
title = {{Llama-3.3-70B-Instruct} Model Card},
year = {2024},
howpublished = {Hugging Face},
url = {https://huggingface.co/meta-llama/Llama-3.3-70B-Instruct},
note = {Model released 2024-12-06; accessed 2026-11-04}
}

@misc{editgenso,
author = {Mehedi Hasan Shanto and Muhammad Asaduzzaman
 and Md Ahsanuzzaman and Alioune Ngom},
title = {Generating Stack Overflow Post Edits with {LLM}s:
 A Taxonomy-Driven Evaluation},
year = {2026},
note = {Accepted Registered Report, 42nd IEEE International
 Conference on Software Maintenance and Evolution
 (ICSME 2026); SSRN preprint},
doi = {10.2139/ssrn.7089478}
}
